\documentclass[alpha-refs]{wiley-article}

\pdfoutput=1

\usepackage{siunitx}
\usepackage{subfigure}

\title{Generating virtual scenarios of multivariate financial data for quantitative trading applications}

\abbrevs{PCA, probabilistic component analysis.}

\author[1]{Javier Franco-Pedroso}
\author[1]{Joaquin Gonzalez-Rodriguez}
\author[2]{Jorge Cubero}
\author[2]{Maria Planas}
\author[2]{Rafael Cobo}
\author[2]{Fernando Pablos}

\affil[1]{Audias Research Group, Escuela Politécnica Superior, Universidad Autónoma de Madrid, Madrid, 28049, Spain}
\affil[2]{ETS Asset Management Factory, Pozuelo de Alarcón, Madrid, 28223, Spain}

\corraddress{Javier Franco-Pedroso, Despacho C-215, Escuela Politécnica Superior, Universidad Autónoma de Madrid, Madrid, 28049, Spain}
\corremail{javier.franco@uam.es}

\fundinginfo{ETS Asset Management Factory, Project Name: ``Tecnologías de tratamiento de señal para mercados financieros''}

\runningauthor{Javier Franco-Pedroso et al.}

\begin{document}

\maketitle

\begin{abstract}

In this paper, we present a novel approach to the generation of virtual scenarios of multivariate financial data of arbitrary length and composition of assets. With this approach, decades of realistic time-synchronized data can be simulated for a large number of assets, producing diverse scenarios to test and improve quantitative investment strategies. Our approach is based on the analysis and synthesis of the time-dependent individual and joint characteristics of real financial time series, using stochastic sequences of market trends to draw multivariate returns from time-dependent probability functions preserving both distributional properties of asset returns and time-dependent correlation among time series. Moreover, new time-synchronized assets can be arbitrarily generated through a PCA-based procedure to obtain any number of assets in the final virtual scenario. For the validation of such simulated data, they are tested with an extensive set of measurements showing a significant degree of agreement with the reference performance of real financial series, better than that obtained with other classical and state-of-the-art approaches.

\end{abstract}

\section{Introduction}

Quantitative analysis and algorithm-based investment nowadays have a relevant presence in stock markets, where a high percentage of trade orders are currently driven by computational finance. Those systems exploit the inefficiencies or inaccuracies of the market through the analysis of previous data and real-time predictions over running markets, detecting and taking advantage of time-and-asset localized opportunities or managing risk in uncertain scenarios by dynamically adapting portfolio weights.\par

Obviously, there is just one single historical record of financial data for a given market. This record is the unique data resource available to develop, optimize and test those algorithmic investment strategies. Once an investment strategy has been developed and tested, we know its performance on some specific period of the historical record, but there is no way to know how the algorithmic strategy would perform in an unseen scenario, different from what has happened in real markets before. It would then be desirable to have a procedure that generates new hypothetical but plausible realizations of financial time series, in order to test the robustness and performance of current quantitative investment strategies, or to develop and optimize new ones. Moreover, this should be done not only for isolated assets, as they are not independent of each other (specially if they belong to the same market), but for sets of assets following a multivariate approach. The aim of this work is to develop a technique that allows to simulate virtual scenarios of given financial markets involving hundreds or thousands of assets for time periods of arbitrary length that could be used for the aforementioned purposes.\par

In order to test any method that attempts to simulate the behavior of real assets, some properties of real financial time series are needed to be defined first, and then checked on the artificial ones. While there is no  unique and clear definition of how a real financial time series should be, we can observe a set of ``... constraints that a stochastic process has to verify in order to reproduce the statistical properties of returns accurately...'' (\cite{cont01}): returns distributions show heavy tails, high-volatility events tend to cluster in time, and price movements do not show significant autocorrelation, among others. Thus, we firstly introduce in Section \ref{sec:assesment} the set of metrics used that attempt to capture the properties of financial time series and markets. The objective of those metrics is not to validate any ``model'', but to validate the artificial series produced according to those sanity-check ``constraints'', whatever its generation process. The results of these measurements will be reported for our benchmark dataset, consisting of a 16-year period of daily stock returns, in order to compare with the simulated scenarios obtained with different methods.\par

Many models that attempt to describe the evolution of asset prices or returns have been proposed, from the most simple geometric Brownian Motion (GBM) and other variants of stochastic differential equations (SDEs), to the Autoregressive Conditional Heteroskedasticity (ARCH) family (\cite{Bollerslev86}) or the stochastic volatility models (\cite{Asai06}) that take into account the time-varying volatility. However, none of these models has been designed for generating virtual scenarios for high-dimensional multivariate financial datasets, failing either to reproduce some specific features of financial time series or being computationally infeasible. In Section \ref{sec:other_approaches} we will give a brief overview of some of the models that can be used for simulating asset returns that can be found in publicly available toolboxes, and some of their weaknesses will be highlighted based on the set of analysis previously described in Section \ref{sec:assesment}.\par

\begin{figure}[h!]
\begin{center}
\begin{minipage}{\textwidth}
\subfigure[]{
\resizebox*{0.5\textwidth}{!}{\includegraphics{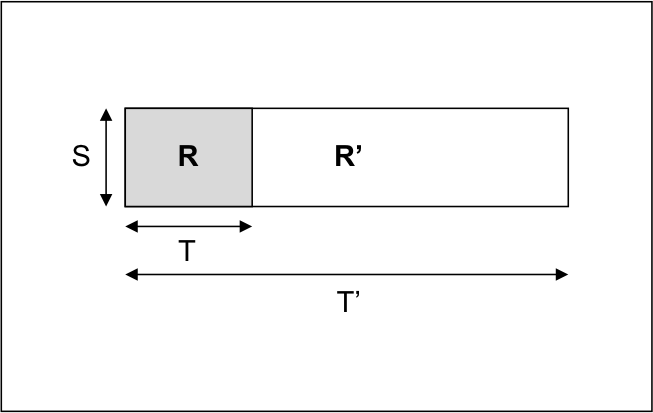}}\label{fig:enlarge}}
\subfigure[]{
\resizebox*{0.5\textwidth}{!}{\includegraphics{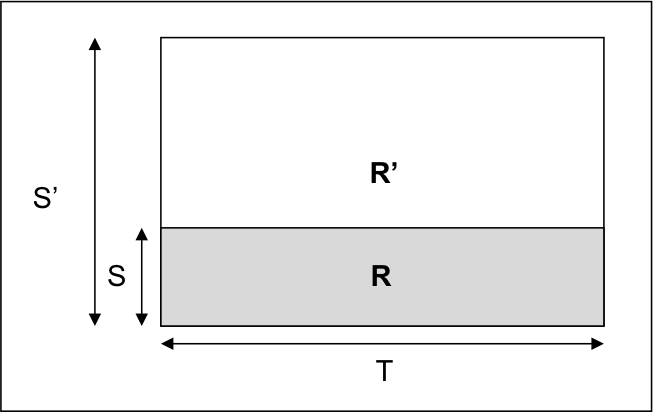}}\label{fig:expand}}
\end{minipage}
\caption{Challenges addressed in this work (\textbf{R} denotes a matrix of T return values from S assets): (a) obtaining a long-term virtual scenario \textbf{R}' (S $\times$ T') from a given dataset \textbf{R} (S $\times$ T), and (b) obtaining a larger dataset \textbf{R}' (S' $\times$ T) from a given dataset \textbf{R} (S $\times$ T).
\label{fig:challenges}}
\end{center}
\end{figure}

For the generation of simulated data that comply with the specified requirements, we propose a synthesis-by-analysis approach, described in Section \ref{sec:approach}, that allows to produce never-seen-before virtual scenarios of a given market as long as desired (Figure \ref{fig:enlarge}). This method is based on segmenting the given dataset in different time periods defined by the market trends, from which time-dependent statistics of the multivariate return distributions are learned. Then, stochastic sequences as long as desired of alternating trends are synthesized by drawing multivariate returns from probability density functions with such time-dependent parameters. In this way, both distributional properties of asset returns and time-dependent correlation among time series are properly reproduced to a great extent.\par

Furthermore, for any investment strategy to be robust, it is desirable that it be tested on unseen assets different from those already observed but presenting a behavior coherent with the market they belong to. In Section \ref{sec:PCA}, a technique for generating new artificial assets that comply with this constraint (figure \ref{fig:expand}) is proposed, allowing to test investment strategies on a much diverse virtual scenario that can involve thousands of assets, including or not the original ones (being in this latter case completely unseen for the investment strategy). To our knowledge, this issue has not been addressed in any previously published study.\par

In Section \ref{sec:long-term}, both proposed techniques are combined in order to generate high-dimensional long-term virtual scenarios of multivariate financial data involving several decades of daily asset returns from  thousands of stocks, whose properties will be checked with our set of metrics. Finally, Section \ref{sec:conclusions} will summarize the conclusions and main contributions of this work.\par

\section{Assessment of simulated financial time series}
\label{sec:assesment}

Several quantitative properties of financial time series have been described in the literature, being known as {\lq}stylized{\rq} or {\lq}empirical{\rq} facts (\cite{cont01,cont07}, \cite{chakraborti07}). Most of them are focused on univariate time series and attempt to describe the distributional properties of asset returns and their dependence properties (\cite{davis99}), while some others try to account for the characteristics of the path defined by price movements. However, when dealing with several assets belonging to the same or different markets, multivariate analysis play a crucial role due to the cross-asset dependencies (\cite{plerou99}). Furthermore, the dependence structure between different assets may be more complex than what can be captured by average correlation coefficients, so additional ways to measure this dependency will be introduced below. All those properties, and the metrics used to quantify them, will be defined in this Section. But first, we will describe the financial dataset used to illustrate those properties and to check the ability of different simulation methods to reproduce their properties.\par

\subsection{Dataset description}
\label{ssec:data}

\begin{figure}[h!]
\begin{center}
\begin{minipage}{\textwidth}
\subfigure[330-stock dataset.]{
\resizebox*{0.5\textwidth}{!}{\includegraphics{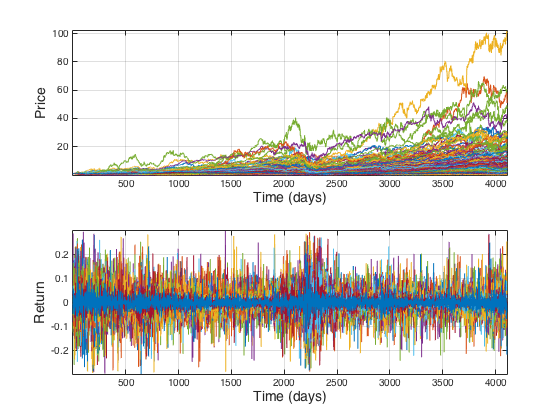}}\label{fig:stocks}}
\subfigure[Equally-weighted market index.]{
\resizebox*{0.5\textwidth}{!}{\includegraphics{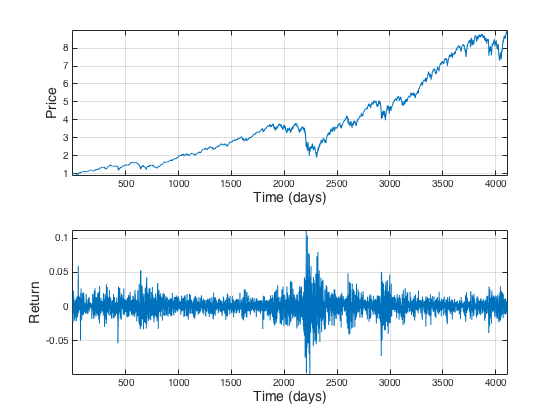}}\label{fig:market}}
\end{minipage}
\caption{Time series of our 330-asset dataset (a), comprising stocks that have been part of the S\&P500 index at some time between 01/01/2000 and 04/29/2016, and its equally-weighted market index (b). For both subfigures, price (upper panel) and return (lower panel) series are shown.}
\label{fig:dataset}
\end{center}
\end{figure}

In this work, we will deal with stocks. Particularly, we will analyze and use the daily prices/returns  between 01/01/2000 and 04/29/2016 of a set of 330 stocks that have been part of the S\&P500 index at some time within this given period. This dataset is illustrated in Figure \ref{fig:stocks}, showing the time series of both prices (upper panel) and returns (lower panel). For a better visualization, stock prices have been forced to start at a price value $p(t = 0) = 1$. This kind of time series representation will be used throughout this paper to illustrate both real and simulated financial time series.\par

On the other hand, an entire dataset belonging to the same asset class can be represented by a market index, an aggregated value produced by combining its constituting stocks or investment vehicles. In this way, the average behavior of the asset class can be easily tracked over time. Also, the resulting time series account for the correlation between different assets, and thus it reflects how these correlations vary over time. In this work, an equally-weighted index is used to represent the overall behavior of a set of stocks within a given time period. Figure \ref{fig:market} shows the equally-weighted market index for our particular dataset.\par

\subsection{Empirical properties of financial time series}
\label{ssec:properties}

\subsubsection{Distributional properties}
\label{sssec:distrib}

Most of the metrics used in order to quantify empirical properties of financial time series describe asset returns as a random variable. In this work, we will consider simple returns, $R_t$. Among the different statistics that can be obtained from return time series, the most used are those related to the shape of their distribution. The distribution of asset returns tends to be non-Gaussian, shape-piked and presenting heavy tails, as it has been observed in various market data. This behavior is usually quantified by measuring the kurtosis of the distribution of $R_t$, which reflects a deviation from the normal distribution. The heavy-tailed distribution of asset returns is revealed by kurtosis values larger than 3, which is the kurtosis value for a univariate normal distribution.\par

\begin{figure}[h!]
\begin{center}
\begin{minipage}{\textwidth}
\subfigure[Return time series.]{
\resizebox*{0.5\textwidth}{!}{\includegraphics{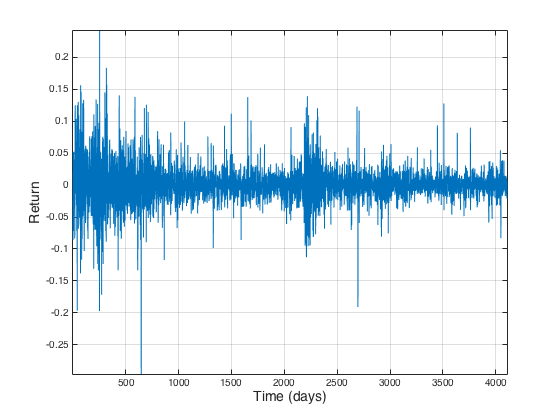}}\label{fig:series}}
\subfigure[Distribution of returns.]{
\resizebox*{0.5\textwidth}{!}{\includegraphics{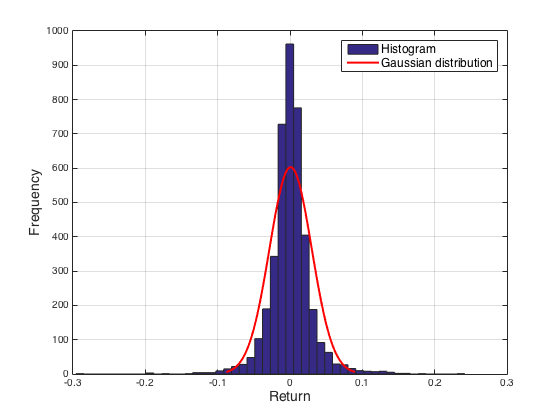}}\label{fig:dist}}
\end{minipage}
\caption{Example of return time series (a) and its distribution (b) for the Adobe Systems Inc. stock. The Gaussian distribution is also plotted for comparison purposes. This particular time series has a kurtosis value of 12.84 and a skewness value of 0.04.}
\label{fig:real_stock}
\end{center}
\end{figure}

\begin{figure}[h!]
\begin{center}
\begin{minipage}{\textwidth}
\subfigure[Kurtosis.]{
\resizebox*{0.5\textwidth}{!}{\includegraphics{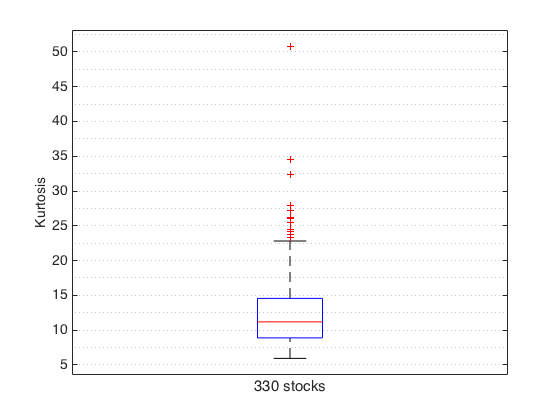}}\label{fig:kurt_dataset}}
\subfigure[Skewness.]{
\resizebox*{0.5\textwidth}{!}{\includegraphics{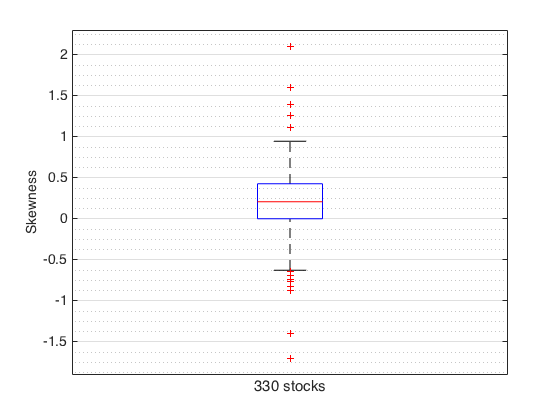}}\label{fig:skew_dataset}}
\end{minipage}
\caption{Box-plots of kurtosis (a) and skewness (b) values in our S\&P500 330-stock dataset.}
\label{fig:kurt_skew_dataset}
\end{center}
\end{figure}

Figure \ref{fig:series} shows the return series for a particular stock within our dataset (Adobe Systems Inc.), while Figure \ref{fig:dist} shows its distribution. As it can be seen, the distribution is more shape-piked and fat-tailed than the normal one, leading to large kurtosis values. This is a general trend in our dataset, as Figure \ref{fig:kurt_dataset} shows, where the distribution of kurtosis values from all 330 stocks is plotted.\par

Also, the distribution of asset returns usually shows a significant degree of asymmetry, which can be quantified by its skewness value. Positive values of skewness reflects more probable values at the right tail of the distribution, while negative values indicates a longer or fatter left tail. For a univariate normal distribution, the skewness value is 0.

Whether the skewness of asset returns is positive or negative seems to depend on the type of financial series being analyzed (\cite{Albuquerque12}): while negative skewness values have been reported for market indexes, presenting larger drawdowns than upward movements, individual stocks seem to show greater variance, tending to positive skewness values. This has been also observed in our dataset: the skewness value for our equally-weighted market index in Figure \ref{fig:market} is -0.0421, while the distribution of skewness values from all 330 stocks, shown in Figure \ref{fig:skew_dataset}, is slightly shifted to positive values (but also presents negative ones).

Distributional properties of asset returns are usually estimated over the whole financial time series, as it has been done above, obtaining a single value of the metric (kurtosis or skewness) per asset. However, these properties may change over time depending on the volatility conditions in different periods, so rolling metrics will also be used in this work in order to account for the time-varying dynamics of the shape of the distribution of asset returns.\par

\subsubsection{Dependence properties}
\label{sssec:dep}

Another well-known fact is that asset returns do not present significant linear autocorrelation, except for very short intra-day time-scales. This fact has been widely documented and has been cited as a support of the {\lq}efficient market hypothesis{\rq} due to its unpredictability. Although autocorrelation of asset returns is insignificant, it has been observed that some non-linear functions of returns, such as absolute or squared returns, do present significant autocorrelation, usually related to the well-known phenomenon of \textit{volatility clustering}. This property can be summarized by saying that {\textquotedblleft}large price variations are more likely to be followed by large price variations{\textquotedblright}.\par

\begin{figure}[h!]
\begin{center}
\begin{minipage}{\textwidth}
\subfigure[ACF of returns.]{
\resizebox*{0.5\textwidth}{!}{\includegraphics{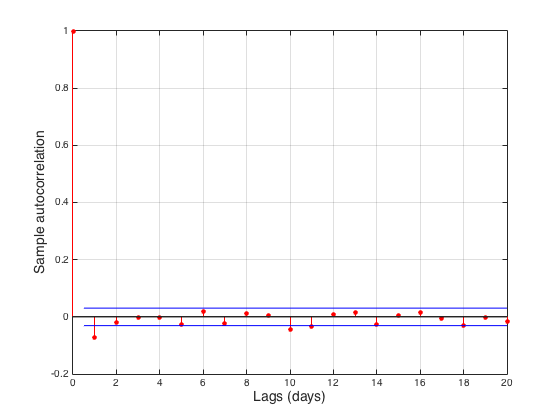}}\label{fig:simples}}
\subfigure[ACF of absolute returns.]{
\resizebox*{0.5\textwidth}{!}{\includegraphics{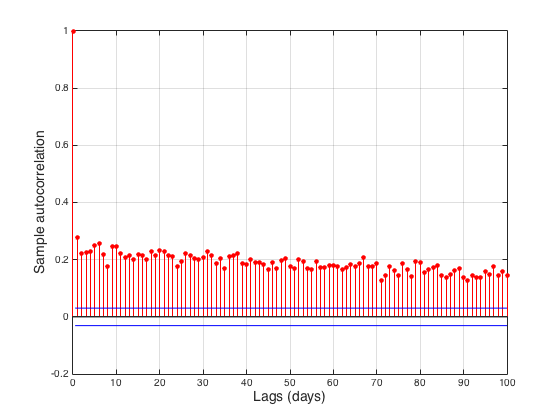}}\label{fig:abs}}
\end{minipage}
\caption{Sample autocorrelation function (ACF) of returns (a) and absolute returns (b) for the Adobe Systems Inc. stock. Confidence bounds ($\pm$0.0312) are also shown.}
\label{fig:autocorrs}
\end{center}
\end{figure}

\begin{figure}[h!]
\begin{center}
\begin{minipage}{\textwidth}
\subfigure[Returns.]{
\resizebox*{0.5\textwidth}{!}{\includegraphics{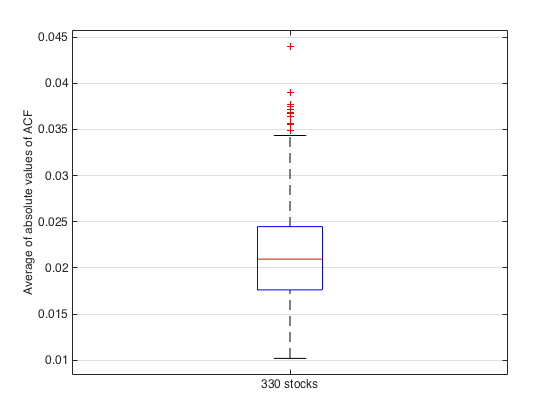}}\label{fig:avg_simples}}
\subfigure[Absolute returns.]{
\resizebox*{0.5\textwidth}{!}{\includegraphics{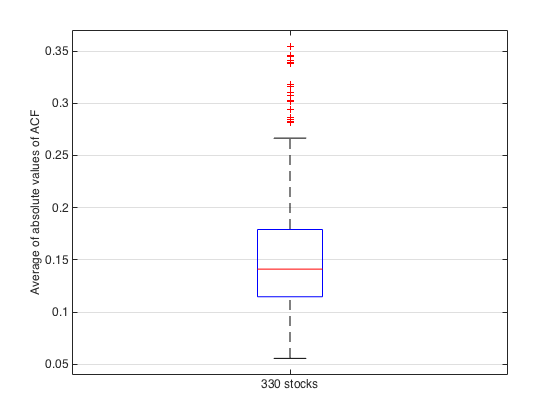}}\label{fig:avg_abs}}
\end{minipage}
\caption{Average of the absolute values of the sample autocorrelation function (ACF) of returns (a) and absolute returns (b) for every stock within our dataset. For returns, the 20 first lags are taken into account, while 100 lags are averaged for absolute returns.}
\label{fig:autocorrs_330}
\end{center}
\end{figure}

Figure \ref{fig:autocorrs} shows the autocorrelation functions (ACF) of returns and absolute returns for the Adobe Systems Inc. stock. As it can be seen, autocorrelation of asset returns rapidly decays to zero (Figure \ref{fig:simples}), while squared returns are positively autocorrelated even for large time-lag values (Figure \ref{fig:abs}). In order to show how rapidly decay autocorrelation values for the whole dataset, the absolute values of the ACF are averaged for every stock (using the 20 first lags for returns and the 100 first lags for absolute returns), and the distribution of the resulting values plotted (see Figure \ref{fig:autocorrs_330}). As shown, the previous observation for a particular stock holds true for the whole dataset.\par

\subsubsection{Pathwise properties}
\label{sssec:path}

Some other properties of univariate financial time series are based on the characteristics of the price time series. For example, the irregularity of the trajectory followed by price movements of a financial asset is related to its risky character. This can be quantified by the number of trends observed over a given time period. However, defining when a trend appears or changes is not straightforward and different definitions can be applied, existing several methods for change-point detection in time series (\cite{Lavielle06}, \cite{Lovric14}). Among other aspects, a noise level or threshold must be defined in order to decide when a given price movement should be considered a trend change or just ``noise'' within the same trend. In this paper, a trend is defined as the price movement between two consecutive directional change (DC) points as defined in \cite{Bakhach16}. However, in this work the goal is not either to summarize the price series nor to predict DC points, but to segment ex-post a given data record for the following purposes:

\begin{itemize}
\item split the historical dataset in the analysis stage of our presented approach, described in Section \ref{ssec:analysis}.

\item compare the number of trends, and the metrics derived from them, observed in both real and simulated datasets.
\end{itemize} 

\begin{figure}[h!]
\begin{center}
\begin{minipage}{\textwidth}
\subfigure[Price series and detected trends.]{
\resizebox*{0.5\textwidth}{!}{\includegraphics{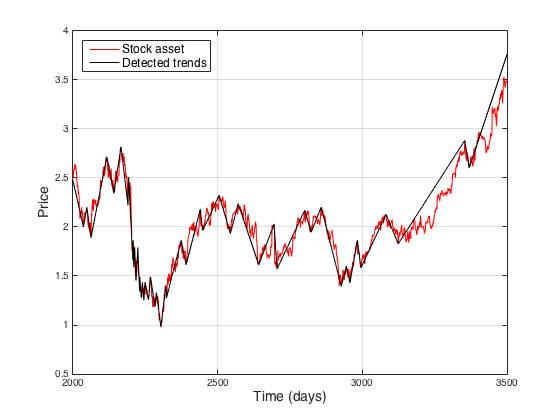}}\label{fig:trends_one_stock}}
\subfigure[Distribution of trend ratios.]{
\resizebox*{0.5\textwidth}{!}{\includegraphics{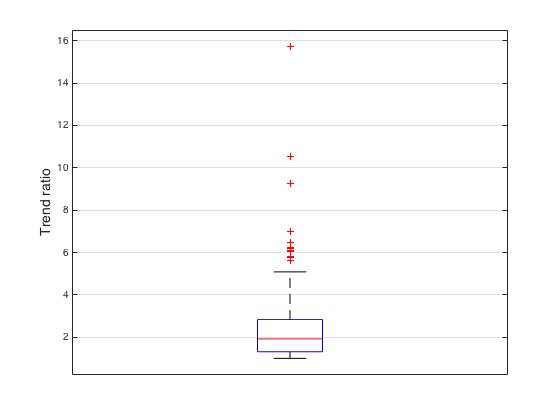}}\label{fig:trends_ratios}}
\end{minipage}
\caption{Detected trends over a 1500-day period of the Adobe Systems Inc. stock (a) and box-plot of the corresponding trend ratios (b).}
\label{fig:trends_and_ratios_one_stock}
\end{center}
\end{figure}

Figure \ref{fig:trends_one_stock} shows an example of the detected trends over a 1500-days period for the Adobe Systems Inc. stock. As it can be seen, a large number of trends can be observed within high volatility periods (between day 2000 and day 2500) while calm periods (between day 3000 and day 3500) present a much lower number of trends.\par

\textit{Trend ratios} can also be used to characterize the behavior of a financial asset. Given a trend, its trend ratio is defined as the return over the period in which the trend is held divided by the noise level used in order to define the trend itself. Very large trend-ratio values will indicate infrequent behavior presenting an unusual return within the given time-period. The distribution of the trend-ratio values for a given time series can then be plotted, revealing high-yield trends as outliers if boxplot-type representations are used. Figure \ref{fig:trends_ratios} show this kind of representation for the Adobe Systems Inc. stock. As it can be seen, some outliers appear due to the high-yield trends in the calm period (between day 3000 and day 3500) shown in Figure \ref{fig:trends_one_stock}.\par

\subsubsection{Cross-asset relationships}
\label{sssec:cross}

\begin{figure}[h!]
\begin{center}
\begin{minipage}{\textwidth}
\begin{center}
\resizebox*{0.5\textwidth}{!}{\includegraphics{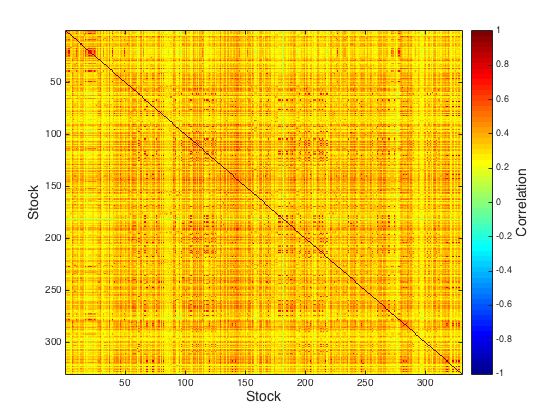}}
\end{center}
\end{minipage}
\caption{Correlation map between our S\&P500 330-stocks dataset.}
\label{fig:corrs_330}
\end{center}
\end{figure}

While it is necessary that individual artificial assets possess univariate characteristics similar to the real ones, this is not sufficient, as investment strategies usually deal with a large number of assets. It is also necessary then that the whole artificial dataset behaves synchronously as a real one, presenting similar cross-asset relationships as those observed in assets from real markets. The most straightforward way to quantify these cross-asset relationships is to compute the correlation coefficient between each pair of assets. An easy way to visualize and compare correlations among different assets and datasets is to obtain the correlation matrix for the dataset (which is symmetric) and graphically represent those coefficients in what it is usually called a \textit{correlation map}. Such kind of representation can be seen in Figure \ref{fig:corrs_330} for our S\&P500 330-stock dataset.\par

\begin{figure}[h!]
\begin{center}
\begin{minipage}{\textwidth}
\subfigure[Low volatility period.]{
\resizebox*{0.5\textwidth}{!}{\includegraphics{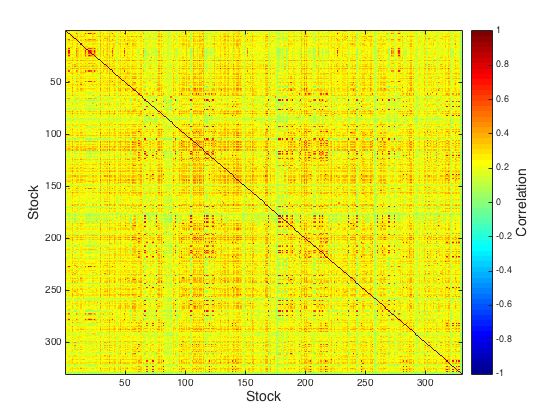}}\label{fig:low_vol}}
\subfigure[High volatility period.]{
\resizebox*{0.5\textwidth}{!}{\includegraphics{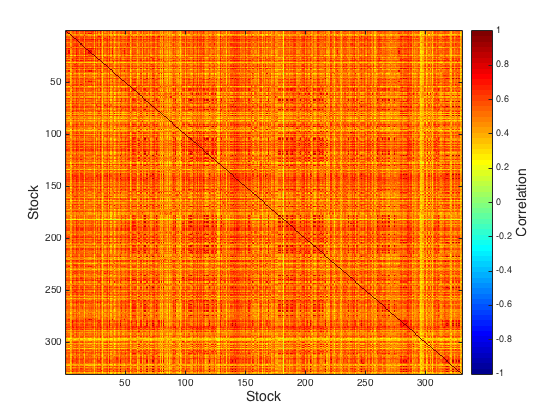}}\label{fig:high_vol}}
\end{minipage}
\caption{Correlation map between our S\&P500 330-stocks dataset two different periods: a low volatility period (a) between days 1000 and 1500 (see time series in Figure \ref{fig:dataset}) and a high volatility period (b) between days 2000 and 2500.}
\label{fig:corrs_vol}
\end{center}
\end{figure}

Although a given market may present correlations between assets that hold on average (for example, companies that belong to the same industry), these correlations usually change dynamically over time. For example, it has been reported that correlations largely increase during high volatility periods, resulting in markets that move as a whole during crisis. This can be seen in Figure \ref{fig:corrs_vol}, where correlation maps of our S\&P500 330-stock dataset have been plotted for two different time periods: a low volatility period (Figure \ref{fig:high_vol}) between days 1000 and 1500 (see time series in Figure \ref{fig:dataset}), and a high volatility period between days 2000 and 2500. As it can be seen, while both correlation maps show similar patterns regarding relative cross-correlation between assets, correlations are much higher during the high volatility period.\par

In order to quantify this dynamic behavior of cross-asset relationships, the equally-weighted market index of the given set of assets will be used to define the following metrics, as this time series depends on the relationship between different assets:\par

\begin{figure}[h!]
\begin{center}
\begin{minipage}{\textwidth}
\subfigure[Temporal evolution.]{
\resizebox*{0.5\textwidth}{!}{\includegraphics{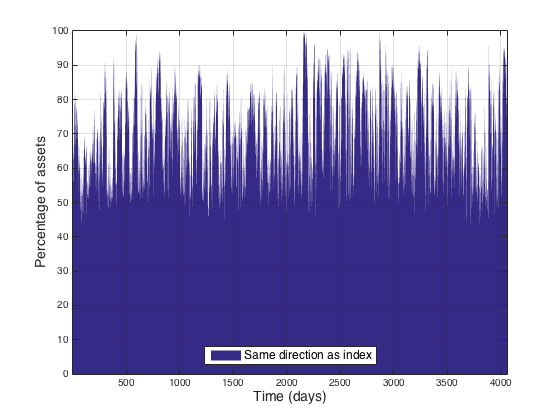}}\label{fig:direc_sim_temp}}
\subfigure[Boxplot distribution.]{
\resizebox*{0.5\textwidth}{!}{\includegraphics{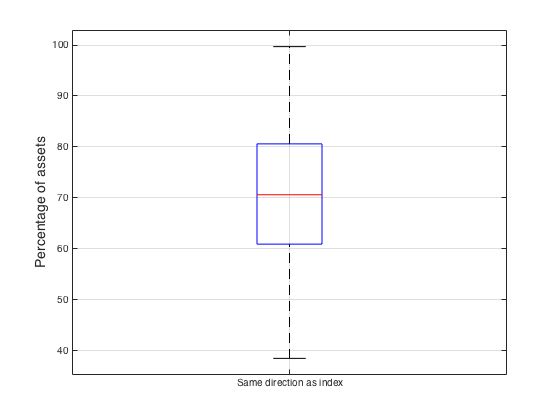}}\label{fig:direc_sim_box}}
\end{minipage}
\caption{Temporal evolution (a) and box-plot (b) of directional similarity values for our equally-weighted 330-stock market.
\label{fig:direc_sim}}
\end{center}
\end{figure}

\begin{itemize}
\item \textit{Number of trends}. We will require that market indexes of artificial assets present a similar behavior to that of the real ones in terms of number of trends over a certain period. If artificial time series would be uncorrelated, their aggregation will result in a smoothed path presenting much lower number of trends.

\item \textit{Directional similarity}. This quantity measures the percentage of assets within a given time period whose prices move in the same direction (either upward or downward) than the equally-weighted market index, reflecting thus the temporal correlation among different financial series. A one-day stepped fifty-day sliding-window is used to obtain the moving average price of the market index, and for each window the percentage of assets whose prices move in the same direction are computed. Figure \ref{fig:direc_sim_temp} shows the value of this metric obtained for each of the different windows, while Figure \ref{fig:direc_sim_box} represent their distribution by means of a boxplot. As shown, it is usual that a high percentage of the assets from a given market become correlated at several time periods, being this the expected behavior for artificial assets.
\end{itemize} 

\section{Evaluation of stochastic models as simulation methods}
\label{sec:other_approaches}

\subsection{Classical stochastic models}

Once we have defined some properties of financial time series, we can proceed to test some approaches for simulating virtual scenarios and their ability to reproduce such properties when generating artificial assets. When looking for publicly available methods to simulate price or return values, the most widely used is the Geometric Brownian Motion (GBM), as it is the core of the well-known Black-Scholes model (\cite{Black73, Merton73}). This method describes the evolution of the price of a stock as a stochastic differential equation (SDE) based on a Wiener process (or Brownian motion), the percentage drift and the percentage volatility, which are taken to be constant and equal to the historical average return and standard deviation, respectively.\par

\begin{figure}[h!]
\begin{center}
\begin{minipage}{\textwidth}
\subfigure[Return time series.]{
\resizebox*{0.5\textwidth}{!}{\includegraphics{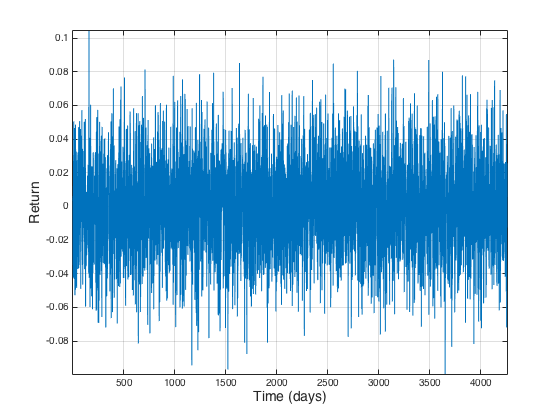}}\label{fig:GBM_series}}
\subfigure[Distribution of returns.]{
\resizebox*{0.5\textwidth}{!}{\includegraphics{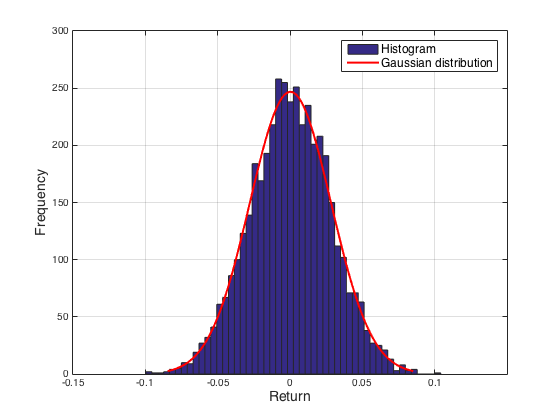}}\label{fig:GBM_dist}}
\end{minipage}
\caption{Simulated return time series (a) and its distribution (b) for a simulated path through GBM. This particular time series has a kurtosis value of 2.98 and a skewness value of -0.03.}
\label{fig:GBM}
\end{center}
\end{figure}

\begin{figure}[h!]
\begin{center}
\begin{minipage}{\textwidth}
\subfigure[ACF of returns.]{
\resizebox*{0.5\textwidth}{!}{\includegraphics{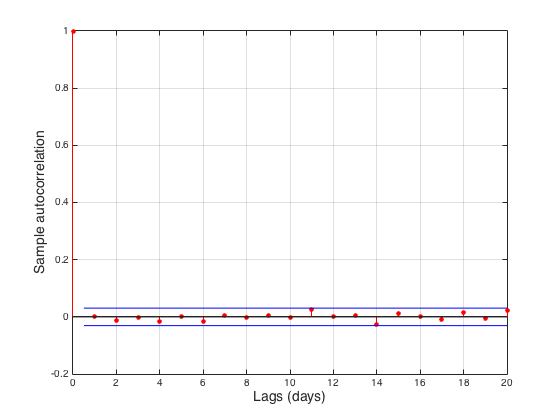}}\label{fig:GBM_simples}}
\subfigure[ACF of absolute returns.]{
\resizebox*{0.5\textwidth}{!}{\includegraphics{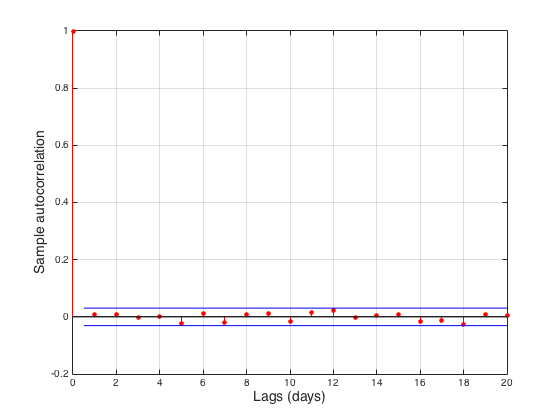}}\label{fig:GBM_abs}}
\end{minipage}
\caption{Example of sample autocorrelation function (ACF) for returns (a) and absolute returns (b) for a simulated path through GBM. Twenty first lags are shown.}
\label{fig:GBM_autocorrs}
\end{center}
\end{figure}

However, it has been shown that this model fails to capture some of the well-known stylized facts, specially those related to the dynamic behavior of markets such as the volatility clustering, but also the excess kurtosis of the unconditional distribution of returns as being based on Gaussian innovations with constant parameters. In order to illustrate this, we have simulated a path for the Adobe Systems Inc. stock by using the MATLAB Financial Toolbox (\cite{matlab}) implementing the GBM. Figure \ref{fig:GBM_series} shows the return time series of the simulated path. As it can be seen in Figure \ref{fig:GBM_dist}, GBM clearly fails to reproduce the excess kurtosis of stock returns, as all the return time series is drawn from the same Gaussian distribution with fixed parameters. For the same reason, the model also fails to reproduce the volatility clustering feature, which arise from observing  sets of samples with time-varying variance. This can be seen by simple visual inspection of Figure \ref{fig:GBM_series}, but also reflected when evaluating the autocorrelation function of absolute returns (see Figure \ref{fig:GBM_abs}) where a lack of autocorrelation is observed even for small values of the time-lag. Similar results are obtained when using other variants based on SDEs.\par

\subsection{Volatility models}

In order to overcome these issues, volatility models were proposed attempting to describe the evolution of asset returns or prices by modelling volatility as a time-varying process. Among them, two main families of volatility models have become the dominant approaches: ARCH (Autoregressive Conditional Heteroskedascity) (\cite{Bollerslev92}) models and stochastic volatility (SV) models. In the former one, the variance is a deterministic value that depends on the $p$ previous values of the time series. Then, analytic expression for the likelihood function of the parameters to be fitted can be derived. Conversely, in stochastic volatility models, volatility is a latent stochastic process, requiring complex techniques (e.g. Markov chain Monte Carlo methods) to compute the likelihood function of the parameters.\par

\begin{figure}[h!]
\begin{center}
\begin{minipage}{\textwidth}
\subfigure[Return time series.]{
\resizebox*{0.5\textwidth}{!}{\includegraphics{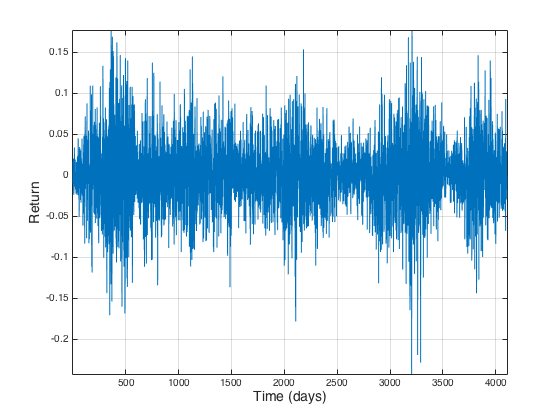}}\label{fig:GARCH_series}}
\subfigure[Distribution of returns.]{
\resizebox*{0.5\textwidth}{!}{\includegraphics{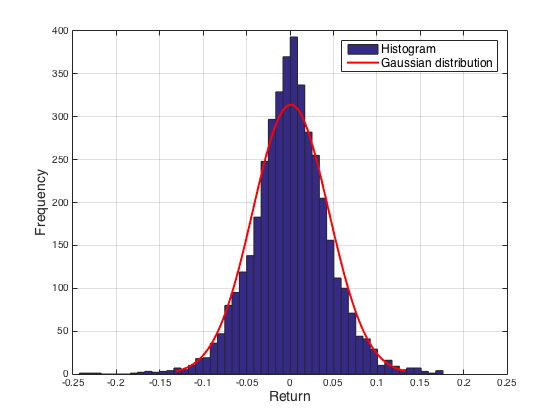}}\label{fig:GARCH_dist}}
\end{minipage}
\caption{Simulated return time series (a) and its distribution (b) for a simulated path through GARCH(1,1). This particular time series has a kurtosis value of 4.44 and a skewness value of -0.06.}
\label{fig:GARCH}
\end{center}
\end{figure}

\begin{figure}[h!]
\begin{center}
\begin{minipage}{\textwidth}
\subfigure[ACF of returns.]{
\resizebox*{0.5\textwidth}{!}{\includegraphics{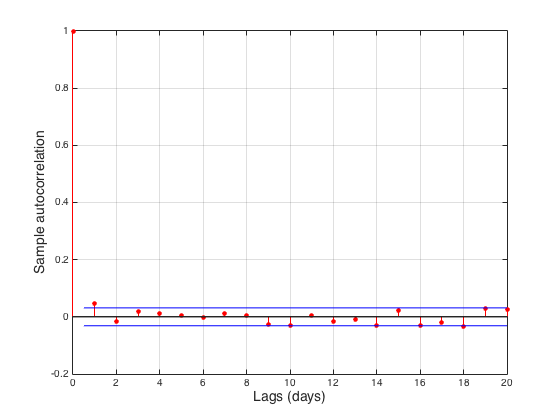}}\label{fig:GARCH_simples}}
\subfigure[ACF of absolute returns.]{
\resizebox*{0.5\textwidth}{!}{\includegraphics{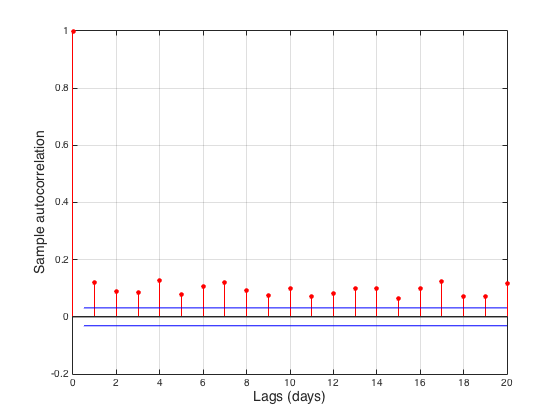}}\label{fig:GARCH_abs}}
\end{minipage}
\caption{Example of sample autocorrelation function (ACF) for returns (a) and absolute returns (b) for a simulated path through GARCH(1,1). Twenty first lags are shown.}
\label{fig:GARCH_autocorrs}
\end{center}
\end{figure}

While both types of models can be used to simulate asset returns or prices, few publicly available implementations can be found that directly provide the simulated paths, as volatility models are mainly used to forecast volatility or to estimate the Value at Risk (VaR) based on those simulations (similarly to how the Black-Scholes model estimate the price of an option based on the simulated paths from a GBM). Most of them are based on the Generalized ARCH (GARCH) model, in which the variance also depends on the $q$ previous values of the variance itself. In the next example, we have used the MATLAB Financial Toolbox to generate a simulated path for a GARCH model with parameters $p=1$ and $q=1$ (GARCH(1,1)) fitted to the Adobe Systems Inc. stock. As it can be seen in Figure \ref{fig:GARCH_series}, the simple visual inspection of the return series reveals that the volatility clustering is reproduced in the simulated path, and this is confirmed by the autocorrelation in absolute returns shown in Figure \ref{fig:GARCH_abs}, while returns are still uncorrelated (Figure \ref{fig:GARCH_simples}). Also, as shown in Figure \ref{fig:GARCH_dist}, the model reproduces some of the excess kurtosis of the return distribution due to this time-varying volatility.\par

\begin{figure}[h!]
\begin{center}
\begin{minipage}{\textwidth}
\subfigure[Distribution of kurtosis values.]{
\resizebox*{0.5\textwidth}{!}{\includegraphics{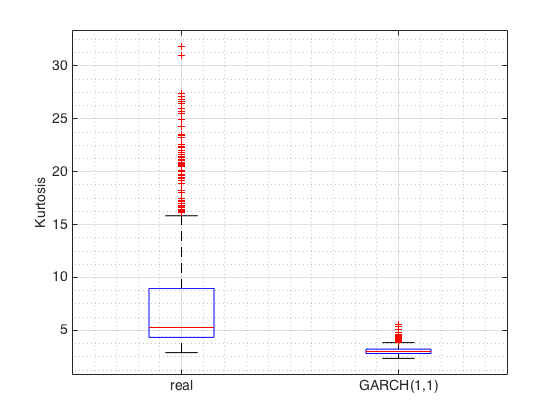}}\label{fig:GARCH_temp_kurt}}
\subfigure[Distribution of skewness values.]{
\resizebox*{0.5\textwidth}{!}{\includegraphics{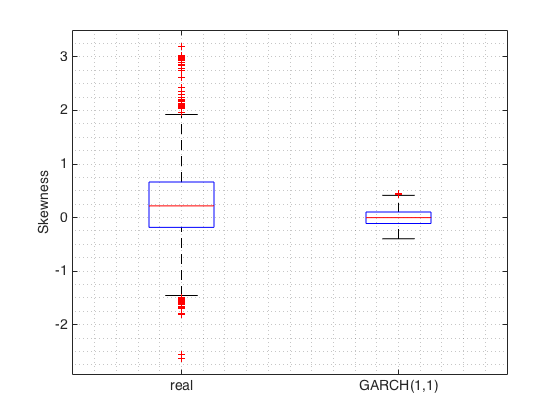}}\label{fig:GARCH_temp_skew}}
\end{minipage}
\caption{Rolling kurtosis (a) and skewness (b) for a simulated path through GARCH(1,1). The temporal analysis is performed through a 180-day sliding-window with 175-day overlapping between consecutive windows.}
\label{fig:GARCH_temp}
\end{center}
\end{figure}

However, much more differences arise, between simulated and real time series, when analysing rolling kurtosis and skewness, as shown in Figure \ref{fig:GARCH_temp}: while for the real time series a wide range of values is observed for both kurtosis and skewness, for the simulated return series most of the values from different time periods lay close to those of a Gaussian distribution (kurtosis = 3 and skewness = 0). This suggests either that the variance in the GARCH(1,1) model vary at a slower rate than in real stocks, or that a different distribution of return innovations is needed.\par

\begin{figure}[h!]
\begin{center}
\begin{minipage}{\textwidth}
\subfigure[Return time series.]{
\resizebox*{0.5\textwidth}{!}{\includegraphics{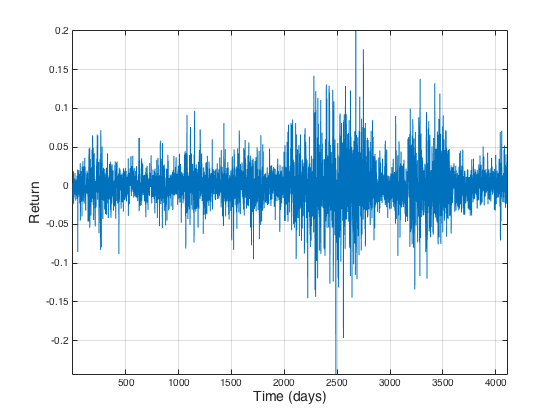}}\label{fig:tGARCH_series}}
\subfigure[Distribution of returns.]{
\resizebox*{0.5\textwidth}{!}{\includegraphics{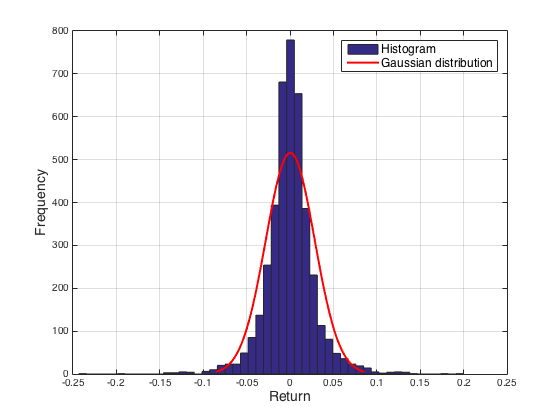}}\label{fig:tGARCH_dist}}
\end{minipage}
\caption{Simulated return time series (a) and its distribution (b) for a simulated path through GARCH(1,1) with Student's t innovations. This particular time series has a kurtosis value of 9.21 and a skewness value of -0.06.}
\label{fig:tGARCH}
\end{center}
\end{figure}

\begin{figure}[h!]
\begin{center}
\begin{minipage}{\textwidth}
\subfigure[Distribution of kurtosis values.]{
\resizebox*{0.5\textwidth}{!}{\includegraphics{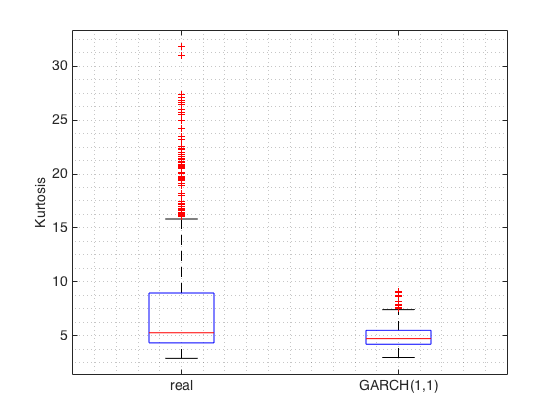}}\label{fig:tGARCH_kurt}}
\subfigure[Distribution of skewness values.]{
\resizebox*{0.5\textwidth}{!}{\includegraphics{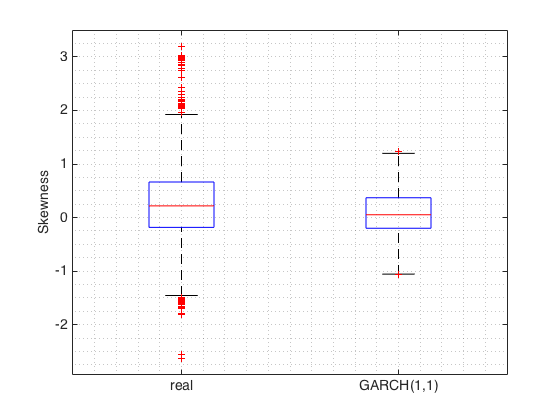}}\label{fig:tGARCH_skew}}
\end{minipage}
\caption{Rolling of kurtosis (a) and skewness (b) for a simulated path through GARCH(1,1) with Student's t innovations. The temporal analysis is performed through a 180-day sliding-window with 175-day overlapping between consecutive windows.}
\label{fig:tGARCH_temp}
\end{center}
\end{figure}

\begin{figure}[h!]
\begin{center}
\begin{minipage}{\textwidth}
\subfigure[Detected trends over the price series (upper panel) and return series (lower panel).]{
\resizebox*{0.5\textwidth}{!}{\includegraphics{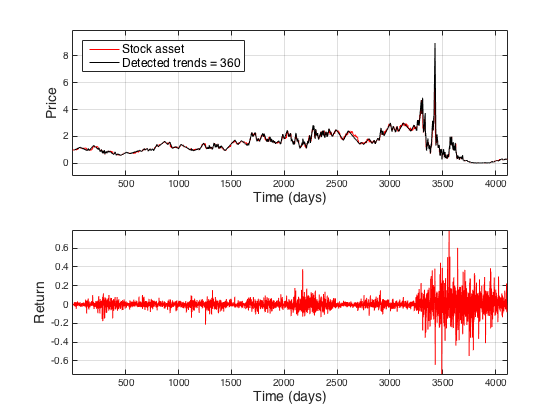}}\label{fig:tGARCH_trends}}
\subfigure[Boxplot of trend ratios.]{
\resizebox*{0.5\textwidth}{!}{\includegraphics{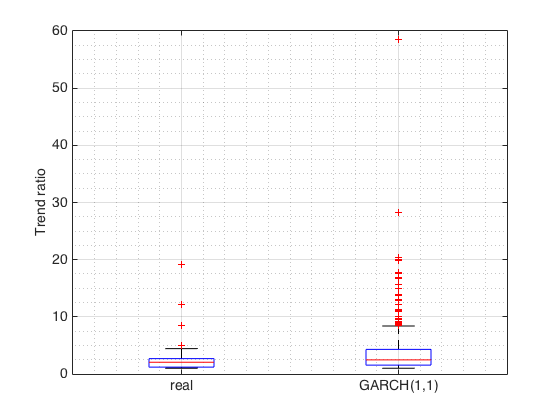}}\label{fig:tGARCH_trend_ratios}}
\end{minipage}
\caption{Detected trends over a simulated path through GARCH(1,1) with Student's t innovations (a) and the distribution of the corresponding trend ratios (b).}
\label{fig:tGARCH_path}
\end{center}
\end{figure}

If Student's \textit{t} innovations are used instead of the Gaussian ones in the GARCH(1,1) model, the excess kurtosis can be better reproduced, as shown in Figure \ref{fig:tGARCH_dist}. For the time-dependent analysis, however, the variance of rolling kurtosis values is still low (see Figure \ref{fig:tGARCH_kurt}), while rolling skewness values vary in a more similar range than that of the real stock (Figure \ref{fig:tGARCH_skew}). Moreover, as only statistical properties are taken into account by GARCH-type models, very rare situations (compared to real financial time series) can appear in simulated time series regarding the pathwise properties. For example, Figure \ref{fig:tGARCH_trends} shows a simulated path with a very high number of trends, some of them presenting very large trend-ratio values, as shown in Figure \ref{fig:tGARCH_trend_ratios}.\par

\subsection{Multivariate volatility models}

Besides the aforementioned issues, multivariate extensions of volatility models (\cite{Bauwens06, Asai06}) present practical difficulties when attempting to fit model parameters to high-dimensional data. Some variants have been developed in order to overcome this issue, but the high computational cost of practical implementations makes them infeasible for large datasets. For example, BEKK (\cite{Engle95}) is a multivariate extension of GARCH in which the dynamic covariance matrix follows an Autoregressive and Moving Average (ARMA) process (\cite{Box70}), with autoregression on $p$ previous variance values and moving average on $q$ previous squared time-series values, being then a generative model that can produce a variety of clustering behavior. However, it is a highly parameterized model, so restricted versions are used in practice in order to mitigate overfitting problems and reduce the computational cost.\par

\begin{figure}[h!]
\begin{center}
\begin{minipage}{\textwidth}
\subfigure[ACF of a stock in a 2-asset model.]{
\resizebox*{0.5\textwidth}{!}{\includegraphics{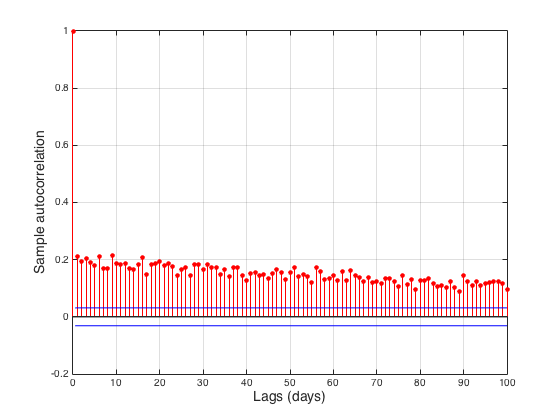}}\label{fig:2_assets}}
\subfigure[ACF of a stock in a 5-asset model.]{
\resizebox*{0.5\textwidth}{!}{\includegraphics{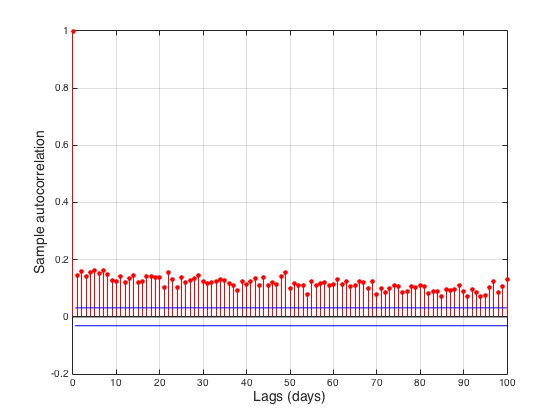}}\label{fig:5_assets}}
\begin{center}
\subfigure[ACF of a stock in a 10-asset model.]{
\resizebox*{0.5\textwidth}{!}{\includegraphics{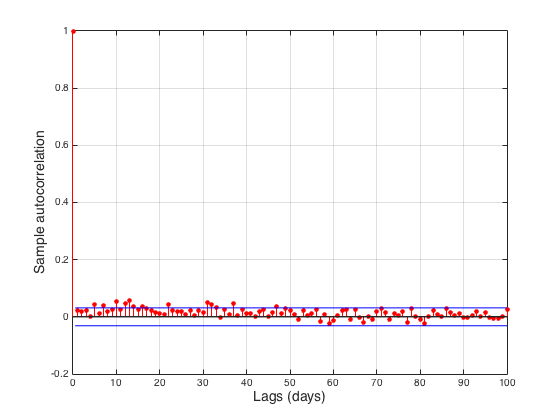}}\label{fig:10_assets}}
\end{center}
\end{minipage}
\end{center}
\caption{Sample autocorrelation of absolute returns for a specific stock when the simulated scenario involves 2 (a), 5 (b) and 10 stocks (c).
\label{fig:BEKK_autocorr}}
\end{figure}

In the following example, the BEKK model implemented in the MFE Toolbox (\cite{MFE}) has been used with parameters $p = 1$ and $q = 1$ to simulate asset returns from multivariate data. Due to the high computational cost of the parameter estimation, the number of modeled assets has been limited to a maximum of 10, randomly selected from our S\&P500 330-stock dataset. The most illustrative example of the curse of dimensionality for such kind of models is shown in Figure \ref{fig:BEKK_autocorr}, in which three different virtual scenarios have been simulated with different number of stocks: 2 (Figure \ref{fig:2_assets}),  5 (Figure \ref{fig:5_assets}) and 10 stocks (Figure \ref{fig:10_assets}). The stock for which the sample autocorrelation of absolute returns is plotted in Figure \ref{fig:BEKK_autocorr} is shared among the three virtual scenarios. As the number of assets in the model is incremented, the volatility clustering is worse reproduced, as reflected by a lower autocorrelation of absolute returns.\par

\section{Generation of virtual scenarios for multivariate data}
\label{sec:approach}

As shown in previous Sections, there are many constraints that artificial financial series must comply with, and not all of them are properly reproduced by the most commonly used stochastic models. Moreover, there is a lack of available procedures to efficiently generate virtual scenarios for high-dimensional multivariate data, as in the usual approaches the fitting process fails to converge even for tens of assets.\par

In this Section we describe an efficient and effective method for generating new artificial asset returns from multivariate data within a given market that allows to reproduce, in a stochastic but somewhat constrained way, the behavior of asset returns observed on real data. As this approach does not attempt to provide an explanation of the returns generation process, latent variables and other parameters affecting the whole generation process can be avoided (and the constraints they require), simplifying then the parameters fitting process.\par

Our proposed approach can be divided into two main stages. In the \textit{analysis} stage (section \ref{ssec:analysis}), the historical record of the development dataset is first segmented into different time periods defined by market trends. Then, multivariate data within each trend is analyzed in order to capture both dynamic and statistical properties of asset return series in that period. In the \textit{synthesis} stage (section \ref{ssec:synthesis}), a random sequence of trends, up to the desired length for the virtual scenario, is first hypothesized. Then, for each trend, learned parameters are recovered and multivariate random asset returns are generated for that trend. Finally, multivariate asset returns from different trends are appended and the virtual scenario for the simulated assets obtained. \par

\subsection{Analysis stage}
\label{ssec:analysis}

\begin{figure}[h!]
\begin{center}
\begin{minipage}{\textwidth}
\subfigure[Detected trends on a historical record.]{
\resizebox*{0.5\textwidth}{!}{\includegraphics{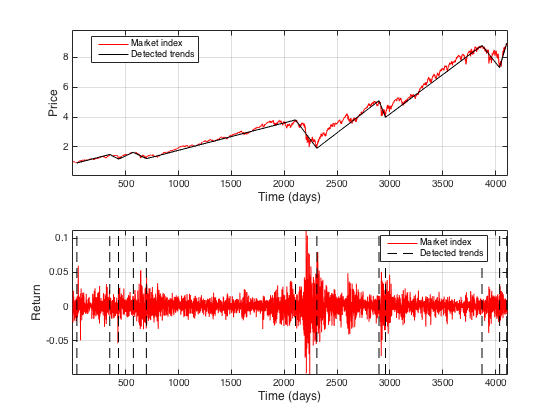}}\label{fig:trends_a}}
\subfigure[Detail of an upward trend.]{
\resizebox*{0.5\textwidth}{!}{\includegraphics{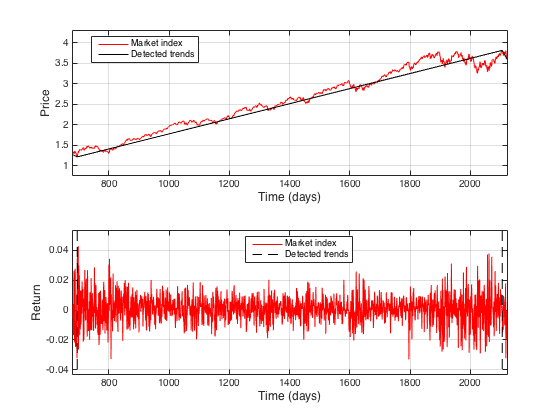}}\label{fig:trends_b}}
\end{minipage}
\caption{Detected trends on an equally-weighted index of a stock market (a) and detail of an upward trend (b).}
\label{fig:trends}
\end{center}
\end{figure}

As it has been previously shown, simulated returns must satisfy not only statistical constraints, but also the path defined by the price movements must show similar trends, avoiding large trend-ratios. However, as being part of a whole market, trends can not be independently modeled for each individual asset, but they must represent an overall behavior. In our approach, an equally-weighted index is used to represent trend changes in the stock market, reflecting the overall behavior of the market within this time period. Based on this market index, the historical record of the multivariate dataset is segmented into different, non-overlapping time periods defined by the detected trends, and the sign of the trend (upward, downward) is recorded. Figure \ref{fig:trends} shows the trends detected over the market index for the dataset described in Section \ref{ssec:data}.\par

Once the historical record has been divided into different trends, time series within a time period corresponding to a trend are analyzed through a short-time sliding window of length $L$ in order to compute the parameters to be reproduced in the synthesis stage. These parameters define the shape of the distribution from which return values, within a sliding window, will be drawn. In this work, a time-dependent multivariate Gaussian distribution is used, whose parameters are assumed to change every $L$ generated samples. That is, for a set of $L$ multivariate asset returns $\mathbf{r}_{t}$ within a window $w$, $\{\mathbf{r}\}_w^L = \{\mathbf{r}_{t-L/2}, \mathbf{r}_{t-L/2+1}, ..., \mathbf{r}_{t-1}, \mathbf{r}_{t}, \mathbf{r}_{t+1}, ..., \mathbf{r}_{t+L/2}\}$, it is assumed that $\mathcal{R}|w \sim \mathcal{N}(\boldsymbol{\mu}_w, \boldsymbol{\Sigma}_w)$, where the mean vector $\boldsymbol{\mu}_w$ and the covariance matrix $\boldsymbol{\Sigma}_w$ are different for every sliding window $w$. It has to be noted that the aim of this Gaussian assumption is not to model the real underlying generation process of asset returns, but to provide an easy way to account for the average price movements of an asset with respect to other assets within the trend, and to capture the dynamic behavior within the trend of variances and correlations. For each given trend, the sequence of parameters obtained from the $W$ sliding windows within the trend, $\{(\boldsymbol{\mu}_1, \boldsymbol{\Sigma}_1)\}$, $\{(\boldsymbol{\mu}_2, \boldsymbol{\Sigma}_2)\}$, ..., $\{(\boldsymbol{\mu}_W, \boldsymbol{\Sigma}_W)\}$, are stored.\par


While this approach resembles a regime-switching model (\cite{Gray96}) in which each type of trend (upward or downward) could be associated with a different regime, note that there are two important differences with our approach: first, the parameters of the generating processes within a trend/regime are also time-varying due to the use of a sliding window much shorter than the trend length; and second, there is no need to estimate transition probabilities between regimes as they are directly observed (in the form of trends) instead of being latent variables.\par

\subsection{Synthesis stage}
\label{ssec:synthesis}

In the synthesis stage, a stochastic sequence of trends is first hypothesized up to the length of the desired simulation, with the only constraint of alternating upward and downward trends. This constraint ensures that the time series do not diverge from the average behavior in the historical record, and that no artificially large trends appear in simulated data, as trends of the same sign cannot be concatenated.\par

Then, for a trend to be simulated, the sequence of $W$ parameters, $\{(\boldsymbol{\mu}_1, \boldsymbol{\Sigma}_1), (\boldsymbol{\mu}_2, \boldsymbol{\Sigma}_2), ..., (\boldsymbol{\mu}_W, \boldsymbol{\Sigma}_W)\}$, within the trend is recovered and $L$ samples drawn from each of the corresponding $W$ multivariate Gaussian distributions. This ensures that volatility clustering is reproduced, as happens with the dynamic behavior of correlations between assets, while new artificial return values are generated.\par

Although the simulated returns are drawn from Gaussian distributions, heavy-tails arise even in the short-term return distribution as it becomes a mixture of multivariate Gaussians whose parameters ($\boldsymbol{\mu}_w, \boldsymbol{\Sigma}_w$) are different for every sliding window $w$, changing every $L$ generated samples. That is, for a given trend, the distribution of simulated asset returns is given by $\mathcal{R} \sim \sum_{w=1}^{W} \mathcal{N}(\boldsymbol{\mu}_w, \boldsymbol{\Sigma}_w)$. Once the return values have been simulated for each trend to be generated, those returns paths are concatenated in order to obtain the final virtual scenario given by the hypothesized trend sequence.\par

\begin{figure}[h!]
\begin{center}
\begin{minipage}{\textwidth}
\subfigure[Return time series.]{
\resizebox*{0.5\textwidth}{!}{\includegraphics{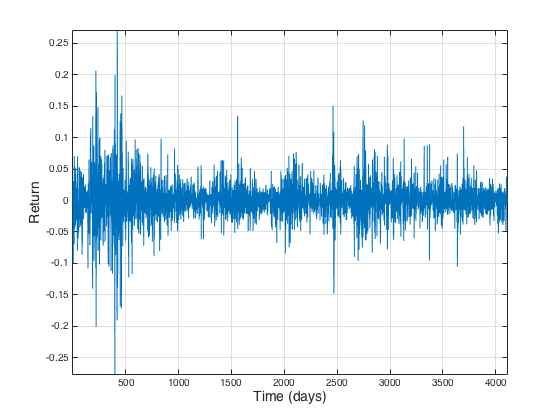}}\label{fig:simulated_stock_returns}}
\subfigure[Return distribution.]{
\resizebox*{0.5\textwidth}{!}{\includegraphics{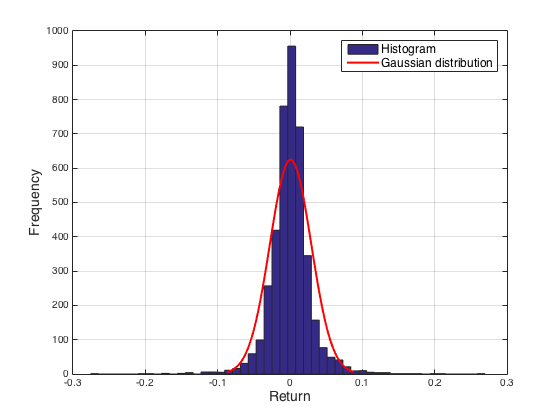}}\label{fig:simulated_stock_dist}}
\end{minipage}
\caption{Example of simulated return time series (a) and its distribution (b) for the Adobe Systems Inc. stock. The Gaussian distribution is also plotted for comparison purposes. This particular time series has a kurtosis value of 13.28 and a skewness value of 0.05.
\label{fig:simulated_stock_1}}\end{center}
\end{figure}
\begin{figure}[h!]
\begin{center}
\begin{minipage}{\textwidth}
\subfigure[Distribution of kurtosis values.]{
\resizebox*{0.5\textwidth}{!}{\includegraphics{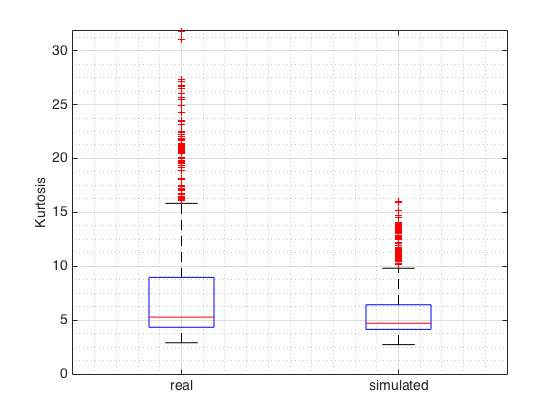}}\label{fig:simulated_stock_kurt_temp}}
\subfigure[Distribution of skewness values.]{
\resizebox*{0.5\textwidth}{!}{\includegraphics{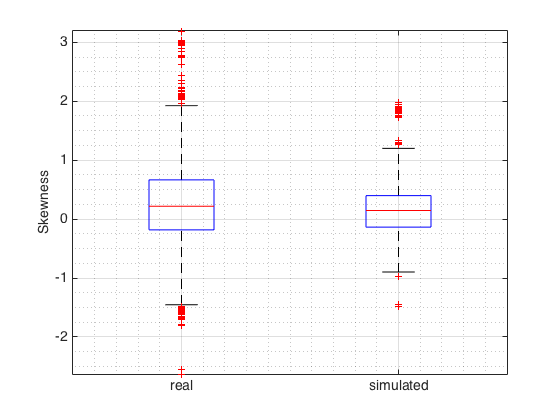}}\label{fig:simulated_stock_skew_temp}}
\end{minipage}
\caption{Rolling kurtosis (a) and skewness (b) for a simulated path of the Adobe Systems Inc. stock. The temporal analysis is performed through a 180-day sliding-window with 175-day overlapping between consecutive windows.
\label{fig:simulated_stock_3}}
\end{center}
\end{figure}

In order to highlight the advantage of the proposed approach over the previously analyzed methods in Section \ref{sec:other_approaches}, we will show some of the properties of the simulated assets in the following Figures. First, the return time series of a particular simulation for the asset in Figure \ref{fig:real_stock} is shown in Figure \ref{fig:simulated_stock_returns} and its corresponding histogram in Figure \ref{fig:simulated_stock_dist}, presenting a leptokurtic distribution similar to that of the real stock (see Figure \ref{fig:dist}). Moreover, when analysing distributional properties over time, it can be seen (Figure \ref{fig:simulated_stock_3}) that rolling kurtosis and skewness values are much closer to the real ones even  compared with simulations obtained with GARCH models with Student's t innovations (see Figure \ref{fig:GARCH_temp}).\par

\begin{figure}[h!]
\begin{center}
\begin{minipage}{\textwidth}
\subfigure[ACF of returns.]{
\resizebox*{0.5\textwidth}{!}{\includegraphics{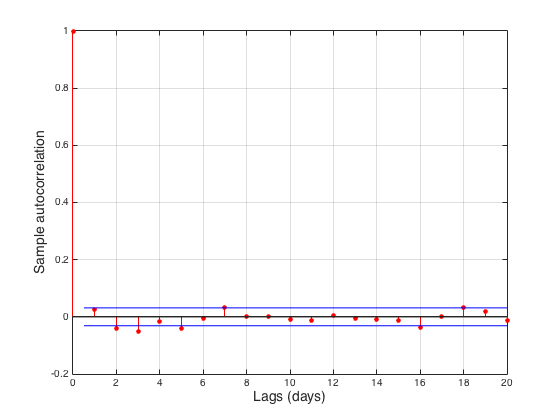}}\label{fig:simulated_stock_simples}}
\subfigure[ACF of absolute returns.]{
\resizebox*{0.5\textwidth}{!}{\includegraphics{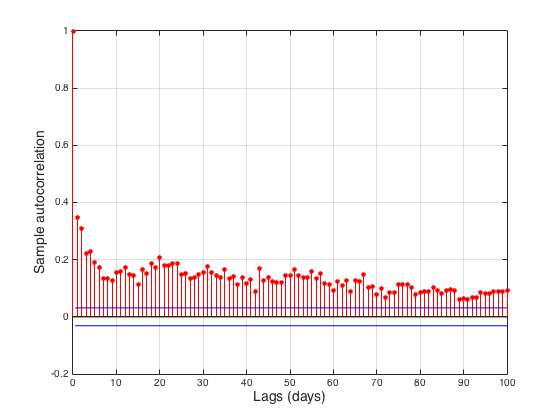}}\label{fig:simulated_stock_abs}}
\end{minipage}
\caption{Sample autocorrelation function (ACF) of returns (a) and absolute returns (b) of a simulated path for the Adobe Systems Inc. stock. Confidence bounds ($\pm$0.0312) are also shown.
\label{fig:simulated_stock_2}}
\end{center}
\end{figure}

Regarding the dependence properties, Figure \ref{fig:simulated_stock_2} shows the ACF of both returns and absolute returns. It is specially remarkable the fact that, although the analyzed asset has been jointly generated with the rest of the 330-stock dataset, absolute returns of the simulated asset present significant autocorrelation for large time-lags (although slightly lower than the real one), while this could not be reproduced with the multivariate GARCH model in Section \ref{sec:other_approaches} (see Figure \ref{fig:BEKK_autocorr}).\par

\begin{figure}[h!]
\begin{center}
\begin{minipage}{\textwidth}
\subfigure[Detected trends on over the price series (upper panel) and return series (lower panel).]{
\resizebox*{0.5\textwidth}{!}{\includegraphics{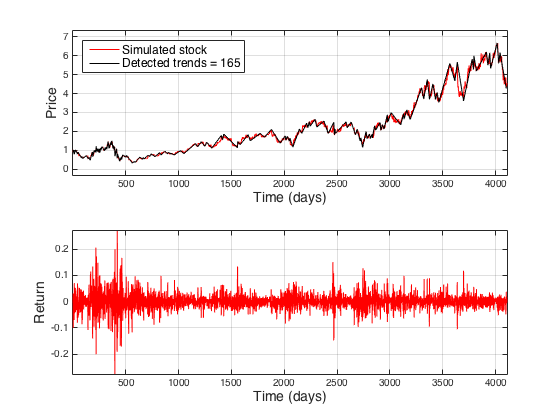}}\label{fig:simulated_stock_trends}}
\subfigure[Distribution of trend ratios.]{
\resizebox*{0.5\textwidth}{!}{\includegraphics{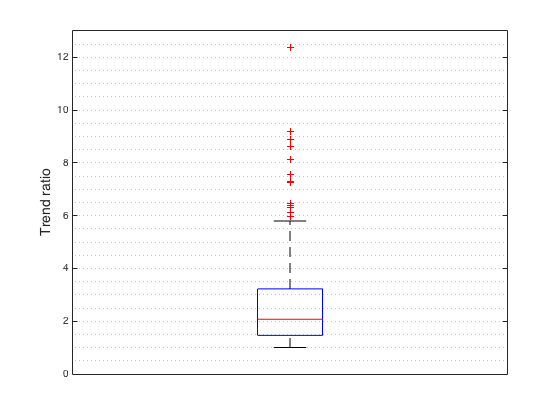}}\label{fig:simulated_stock_trend_ratio}}
\end{minipage}
\caption{Detected trends over the price series (a) and the distribution of the corresponding trend ratios (b) for a simulated path of the Adobe Systems Inc. stock.
\label{fig:simulated_stock_4}}
\end{center}
\end{figure}

Regarding the pathwise properties, Figure \ref{fig:simulated_stock_trends} show the detected trends on the simulated asset, where none artificial trends appear (as those observed in some GARCH simulations, see Figure \ref{fig:tGARCH_trends}). Then, trend-ratios distribution shown in Figure \ref{fig:simulated_stock_trend_ratio} is more similar to that of the real one (see Figure \ref{fig:trends_ratios}), and do not present outliers as those seen in Figure \ref{fig:tGARCH_trend_ratios}.\par

\begin{figure}[h!]
\begin{center}
\begin{minipage}{\textwidth}
\subfigure[Correlation map for the simulated dataset.]{
\resizebox*{0.5\textwidth}{!}{\includegraphics{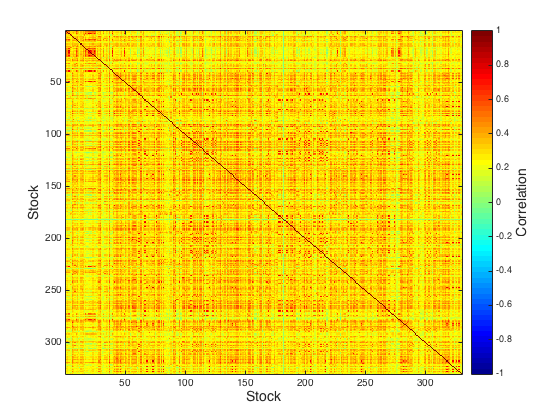}}\label{fig:sim_corrs_330}}
\subfigure[Directional similarity of the simulated dataset.]{
\resizebox*{0.5\textwidth}{!}{\includegraphics{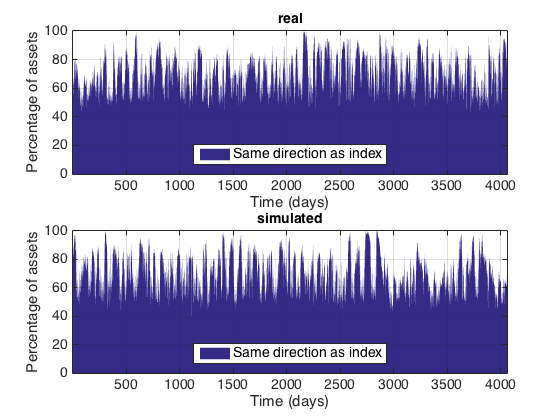}}\label{fig:sim_disim_330}}
\end{minipage}
\caption{Correlation map (a) and directional similarity (b) for the simulated dataset.
\label{fig:virtual_scenarios_corrs}}
\end{center}
\end{figure}

Finally, regarding cross-asset relationships, as the returns in virtual scenarios are drawn from multivariate distributions, cross-correlations between assets are properly reproduced due to the use of the covariance matrices estimated in the analysis step, as it can be seen in the average correlation map shown in Figure \ref{fig:sim_corrs_330}. For the same reason, these cross-correlations change over time, as covariance matrices are updated throughout each trend. This effect is observed in the temporal evolution of directional similarity over time, as shown in Figure \ref{fig:sim_disim_330}.\par

\section{Generation of new artificial assets}
\label{sec:PCA}

Additionally to generating arbitrary long virtual scenarios for a given set of real assets, it would be desirable to test the investment strategies on a much diverse dataset presenting new additional assets with unseen characteristics, but coherent in time with the existing ones (as shown in Figure \ref{fig:expand}). In this way, investment strategies could be fitted to an average market behavior, but not to specific assets, being then more general-purpose strategies. Also, it allows to develop a strategy on a given dataset, but test it in an unseen, different one. In this Section, a method is proposed to synthesize new assets, from a given dataset, by means of a PCA (\cite{bishop06}) projection-recovery process.\par

Principal Component Analysis (PCA) is a technique that allows to represent a set of possibly correlated observations as a linear combination of uncorrelated variables by applying an orthogonal transformation. Some of those uncorrelated variables, the \textit{principal components}, account for most of the variability present in the original data, while the remaining ones can be seen as a source of additive noise for the original variables. Thus, PCA is mainly used as a dimension reduction technique by keeping only the $k$ principal components and discarding the remaining ones. However, in this work PCA will be used with the opposite purpose.\par

When applying PCA to financial time series (\cite{Laloux00}), the first component can be seen as the average behavior of the different time series, in a similar way as a market index does. On the other hand, further components will account for additional variability around this average behavior that explain the differences among the different assets. Each combination of the components, given by the eigenvectors, produces a different asset in the original variable space. Thus, if new artificial eigenvectors are generated based on the original ones, these new combinations of the components will produce new assets different from those in the original dataset, but presenting similar properties as being generated from the same components.\par

\begin{figure}[h!]
\begin{center}
\begin{minipage}{\textwidth}
\subfigure[Direct projection with original transformation matrix.]{
\resizebox*{0.5\textwidth}{!}{\includegraphics[width=1\linewidth]{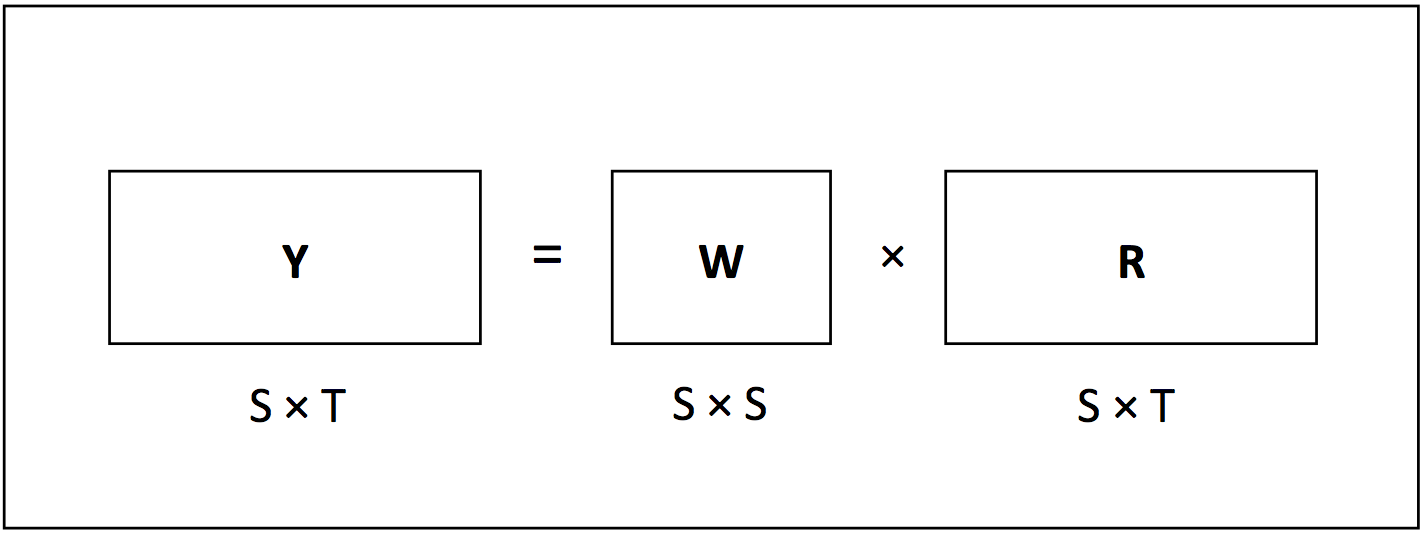}}\label{fig:PCA_direct}}
\subfigure[Inverse projection with artificial transformation matrix.]{
\resizebox*{0.5\textwidth}{!}{\includegraphics[width=1\linewidth]{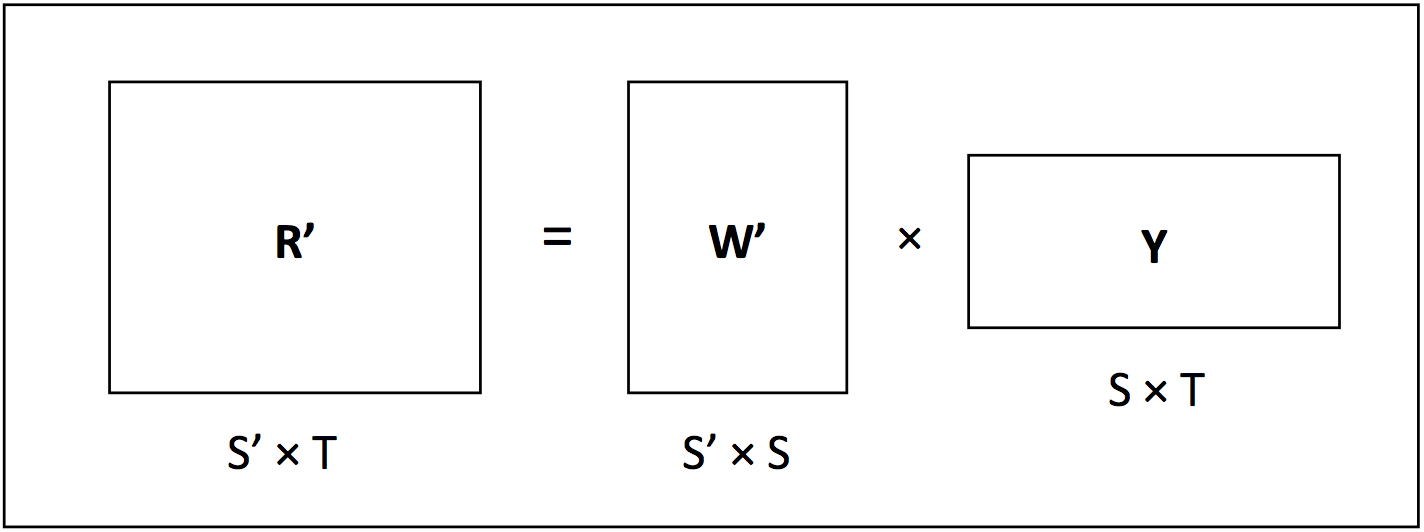}}\label{fig:PCA_inverse}}
\end{minipage}
\caption{Direct (a) and inverse (b) PCA projections in order to obtain new artificial assets.}
\label{fig:PCA}
\end{center}
\end{figure}

For a given dataset $\mathbf{R}$ of $S$ asset returns from $T$ trading days ($S \times T$), a PCA transformation matrix $\mathbf{W}$ can be obtained and these assets projected through $\mathbf{Y} = \mathbf{W}\mathbf{R}$ (Figure \ref{fig:PCA_direct}), being $\mathbf{Y}$ the projected asset returns or components. If no dimension reduction is applied, $\mathbf{W}$ is a $S \times S$ transformation matrix of eigenvectors $\mathbf{w}_s$ that maps the original vectors of asset returns $\mathbf{r}_t$ into their components $\mathbf{y}_t$. For this work, the number of dimensions is not reduced as our goal is to retain the highest degree of variability in order to produce the more diverse virtual scenarios as possible.\par

In order to generate new artificial assets, additional eigenvectors $\mathbf{w'}_s$ can be obtained by some procedure and stacked to form an artificial transformation matrix $\mathbf{W'}$. Then, the components $\mathbf{Y}$ are projected back to the original space by applying the inverse transformation $\mathbf{W'}\mathbf{Y} = \mathbf{R'}$ (Figure \ref{fig:PCA_inverse}), being $\mathbf{R'}$ a new artificially generated set of assets. The procedure used to generate those additional eigenvectors is to draw them from a multivariate normal distribution with parameters $\mathbf{\mu} = E[\mathbf{w}_s]$ and $\mathbf{\Sigma} = Cov(\mathbf{w}_i,\mathbf{w}_j)$. In this way, the average behavior of the artificial assets is close to that of the original dataset and individual artificial assets presents a degree of variability similar to that observed in the original dataset.\par

\begin{figure}[h!]
\begin{center}
\begin{minipage}{\textwidth}
\subfigure[Real assets.]{
\resizebox*{0.5\textwidth}{!}{\includegraphics{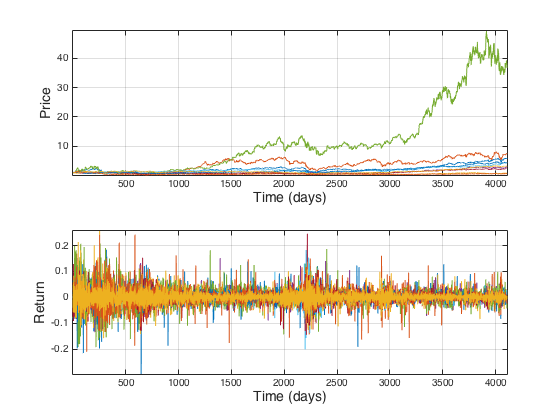}}\label{fig:10_real_assets_PCA}}
\subfigure[Artificial assets.]{
\resizebox*{0.5\textwidth}{!}{\includegraphics{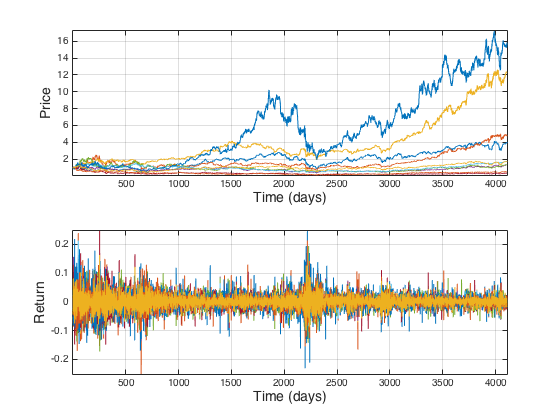}}\label{fig:10_synthetic_assets_PCA}}
\subfigure[Equally-weighted market index of real assets.]{
\resizebox*{0.5\textwidth}{!}{\includegraphics{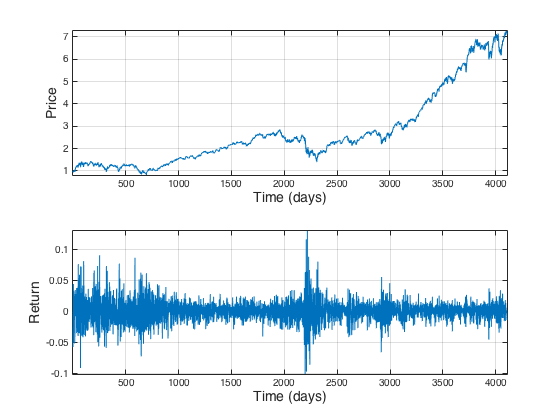}}\label{fig:10_real_assets_PCA_market_index}}
\subfigure[Equally-weighted market index of artificial assets.]{
\resizebox*{0.5\textwidth}{!}{\includegraphics{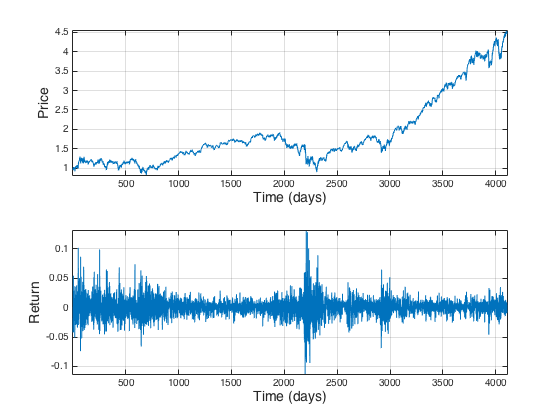}}\label{fig:10_synthetic_assets_PCA_market_index}}
\end{minipage}
\caption{Subset of 10 real (a) and artificial (b) assets, and their corresponding market indexes (c and d, respectively).}
\label{fig:10_assets_PCA}
\end{center}
\end{figure}

Figure \ref{fig:10_assets_PCA} shows an example of the PCA projection-recovery process for a 10-asset subset from our 330-stock dataset, from which a new artificial 10-stock dataset is generated through the described PCA projection-recovery process. As shown, while individual price movements differ from those observed in the real subset (Figure \ref{fig:10_real_assets_PCA}), artificial asset returns (Figure \ref{fig:10_synthetic_assets_PCA}) present a very similar average volatility pattern that follows the average behavior of the market. This is also revealed when plotting the equally-weighted market index of both real and artificial datasets, as shown in Figures \ref{fig:10_real_assets_PCA_market_index} and \ref{fig:10_synthetic_assets_PCA_market_index}, respectively.\par

\begin{figure}[h!]
\begin{center}
\begin{minipage}{\textwidth}
\subfigure[Artificial dataset.]{
\resizebox*{0.5\textwidth}{!}{\includegraphics{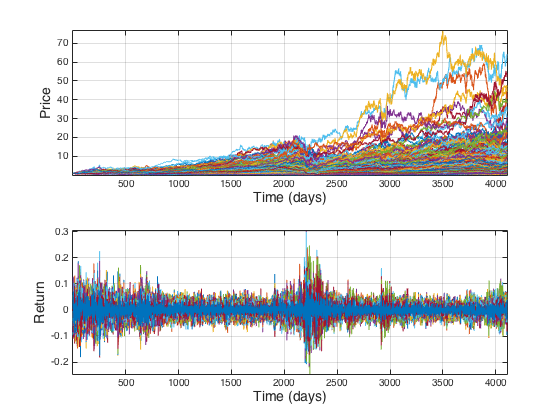}}\label{fig:330_synthetic_assets_PCA}}
\subfigure[Average correlation map.]{
\resizebox*{0.5\textwidth}{!}{\includegraphics{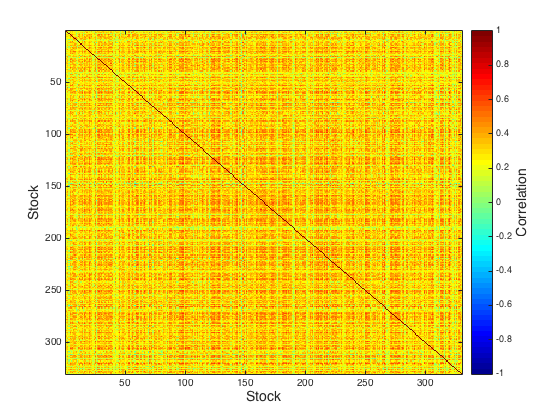}}\label{fig:330_synthetic_assets_PCA_-_corrs}}
\subfigure[Market indexes and detected trends.]{
\resizebox*{0.5\textwidth}{!}{\includegraphics{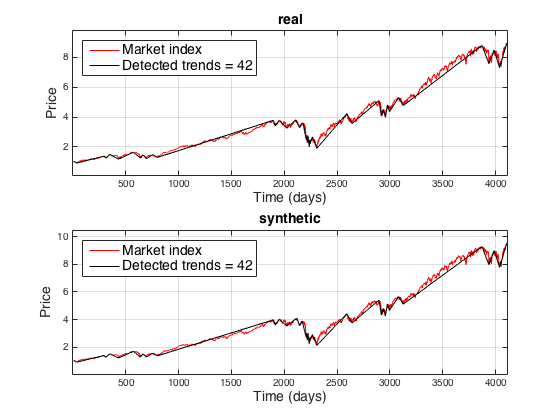}}\label{fig:330_synthetic_assets_PCA_-_trends}}
\subfigure[Distribution of trend ratios.]{
\resizebox*{0.5\textwidth}{!}{\includegraphics{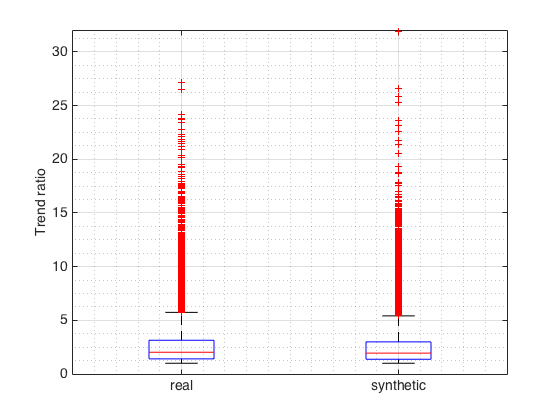}}\label{fig:330_synthetic_assets_PCA_-_trend_ratio}}
\end{minipage}
\caption{Artificial 330-stock dataset (a), its average correlation map (b), detected trends of the equally-weighted market index of both real and artificial 330-stock datasets (c), and trend ratios distributions of individual time series for both real and artificial 330-stock datasets (d).
\label{fig:PCA_330}}
\end{center}
\end{figure}

This average behavior is also reproduced even for large artificial datasets. For example, Figure \ref{fig:PCA_330} shows an artificial dataset (Figure \ref{fig:330_synthetic_assets_PCA}), generated through the described PCA projection-recovery process, of the same size of the original one (330-stock), along with some metrics (Figures \ref{fig:330_synthetic_assets_PCA_-_corrs}, \ref{fig:330_synthetic_assets_PCA_-_trends} and \ref{fig:330_synthetic_assets_PCA_-_trend_ratio}) that reflect a very similar average behavior.\par

\section{Analysis of long-term high-dimensional virtual scenarios}
\label{sec:long-term}

Finally, by combining both previously described techniques, namely the generation of virtual scenarios (section \ref{sec:approach}) and the generation of new artificial assets (section \ref{sec:PCA}), we are able to generate virtual scenarios for large datasets of multivariate data as long and wide as desired. In this way, we can provide large amounts of data to the investment strategies for testing purposes, checking their behavior on unseen scenarios. In this Section, we will look at some generated long-term high-dimensional virtual scenarios and check their properties.\par

\subsection{Experimental framework}
\label{ssec:framework}

The generation of long-term high-dimensional virtual scenarios is based on the real dataset described in Section \ref{ssec:data}, composed of the daily prices/returns from 330 stocks that have been part of the S\&P500 index at some time between 01/01/2000 and 04/29/2016. From this period, trading days in which returns are zero for every asset have been removed, as it is assumed that the market is closed, resulting in 4108-day time series. Table \ref{tab:datasets} shows the size of datasets at different stages of our proposed approach. First, by following the procedure described in Section \ref{ssec:analysis}, trends are detected on the equally-weighted market index of the real dataset, and the time-varying parameters of a Gaussian multivariate distribution estimated for each of those trends (second column in Table \ref{tab:datasets}). Then, the procedure described in Section \ref{ssec:synthesis} is applied to simulate a 50-year virtual scenario for the original set of 330 stocks, resulting in a dataset composed of 12500 trading days from 330 stocks, as shown in the third column of Table \ref{tab:datasets}. Finally, the number of assets is increased up to 1500 by using the PCA projection-recovery process described in \ref{sec:PCA} (fourth column in Table \ref{tab:datasets}).\par

\begin{table}[h!]
\caption{Size of datasets at different stages of the proposed approach.}
\begin{threeparttable}
\begin{tabular}{@{}lccc}
\headrow
\thead{Stage} & \thead{Step 0 (analysis)} & \thead{Step 1 (synthesis)} & \thead{Step 2 (PCA)}\\
  \textbf{Size} (\#days/\#assets) & 4108/330$^{\rm a}$ & 12500/330$^{\rm b}$ & 12500/1500$^{\rm b}$  \\
\hline  
\end{tabular}

\begin{tablenotes}
\item{$^{\rm a}$Real dataset.}
\item{$^{\rm b}$Virtual scenario.}
\end{tablenotes}
\end{threeparttable}
\label{tab:datasets}
\end{table}

In order to show the properties of the generated datasets on a variety of scenarios, three different simulations have been generated following the process described above. As both the sequence of trends and the multivariate returns drawn within each window are randomly generated, different datasets will show different historical market indexes. These overall behaviors will be compared with a 50-year history of the S\&P 500 index between 12/31/1963 and 06/12/2015. On the other hand, it is expected that the different virtual scenarios will show empirical properties similar to those measured on the real dataset from which they have been generated, so we will compare with this dataset the results of our set of metrics.\par

\subsection{Results}
\label{ssec:results}

\begin{figure}[h!]
\begin{center}
\begin{minipage}{\textwidth}
\subfigure[S\&P 500 (95 trends).]{
\resizebox*{0.5\textwidth}{!}{\includegraphics{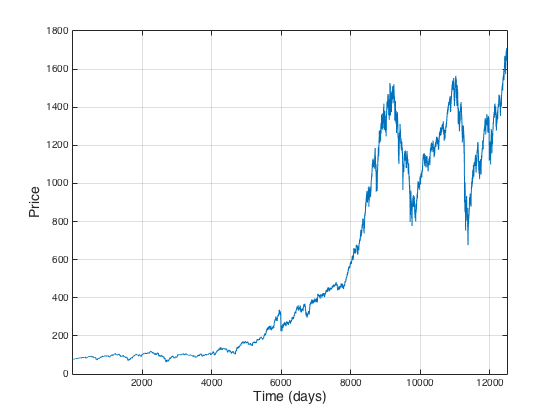}}\label{fig:sp500}}
\subfigure[Virtual scenario 1 (127 trends).]{
\resizebox*{0.5\textwidth}{!}{\includegraphics{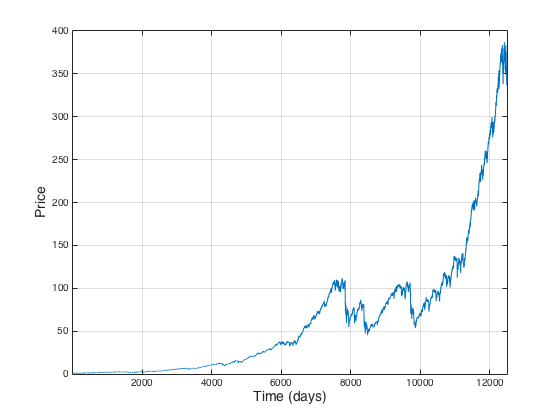}}\label{fig:sim1}}
\subfigure[Virtual scenario 2 (182 trends).]{
\resizebox*{0.5\textwidth}{!}{\includegraphics{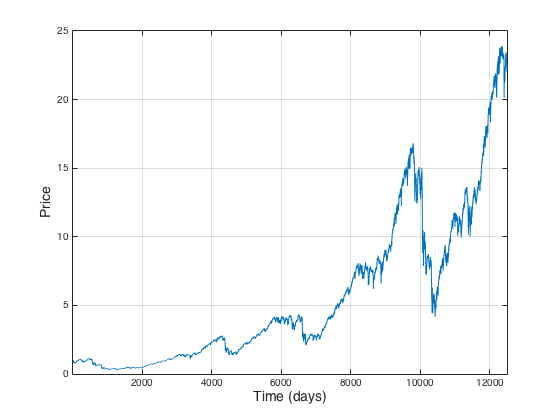}}\label{fig:sim2}}
\subfigure[Virtual scenario 3 (210 trends).]{
\resizebox*{0.5\textwidth}{!}{\includegraphics{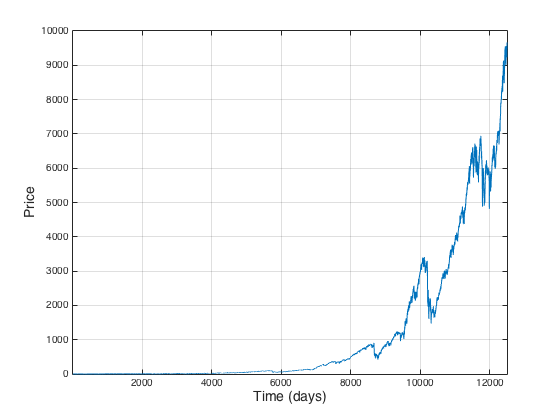}}\label{fig:sim3}}
\end{minipage}
\caption{Price series for a 50-year history of the S\&P 500 index between 12/31/1963 and 06/12/2015 (a) and for the market indexes of three different 50-year 1500-stock virtual scenarios (b, c and d).
\label{fig:50_year_market_indices}}
\end{center}
\end{figure}

Figure \ref{fig:50_year_market_indices} shows the market indexes of three 50-year virtual scenarios of a market involving 1500 assets generated from our 330-stock original dataset, along with a 50-year history of the S\&P 500 index between 12/31/1963 and 06/12/2015. As shown, generated scenarios are highly variable in terms of the aggregated effect reflected in the market index, while presenting an upward pattern similar to that of a real market index. The number of detected trends on the market index for the virtual scenarios is of the order of hundreds, slightly higher to that of our real market index depending on the particular simulation. It should be noted, however, that the real S\&P 500 index is the aggregation of different stocks, not equally weighted, that may change over time (consist of different sets of assets at different time periods), while the market index of our virtual scenarios consist of the same set of equally-weighted assets along its whole history. Also, the number of assets in virtual scenarios is much higher.\par

\begin{figure}[h!]
\begin{center}
\begin{minipage}{\textwidth}
\subfigure[Kurtosis.]{
\resizebox*{0.5\textwidth}{!}{\includegraphics{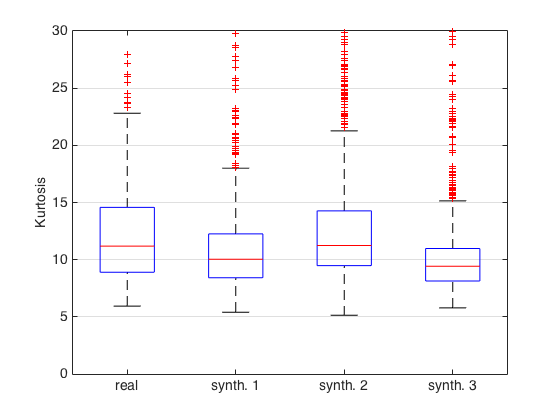}}\label{fig:kurt_sims}}
\subfigure[Skewness.]{
\resizebox*{0.5\textwidth}{!}{\includegraphics{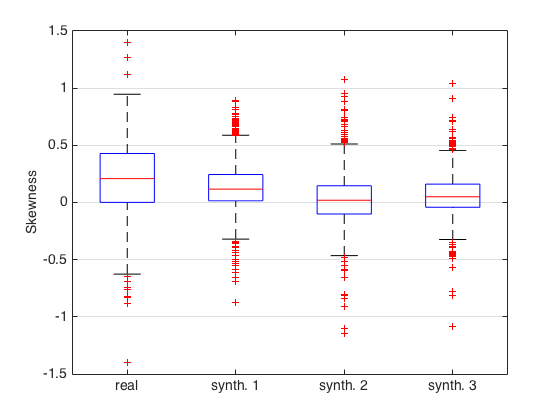}}\label{fig:skew_sims}}
\subfigure[Rolling kurtosis.]{
\resizebox*{0.5\textwidth}{!}{\includegraphics{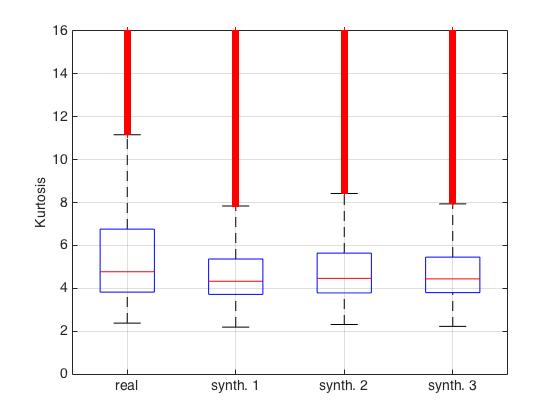}}\label{fig:rol_kurt_sims}}
\subfigure[Rolling skewness.]{
\resizebox*{0.5\textwidth}{!}{\includegraphics{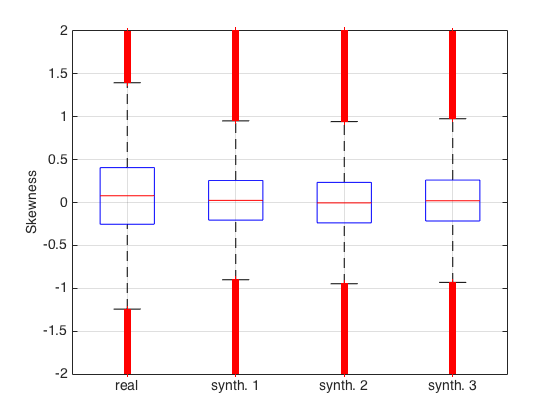}}\label{fig:rol_skew_sims}}
\end{minipage}
\caption{Distributions of kurtosis (a, c) and skewness (b, d) values for our 16-year 330-stock dataset and for three 50-year 1500-stock virtual scenarios. Metrics are computed for the whole time series (a, b) and in a time-varying fashion (c, d).
\label{fig:kurtosis_skewness_sims}}
\end{center}
\end{figure}

In Figure \ref{fig:kurtosis_skewness_sims} we analyze some distributional properties of our three virtual scenarios and compare them with those of our 16-year 330-stock original dataset. Figures \ref{fig:kurt_sims} and \ref{fig:skew_sims} show the distribution of kurtosis and skewness values, respectively, when computed for each whole time series, while Figures \ref{fig:rol_kurt_sims} and \ref{fig:rol_skew_sims} show the same distributional properties but analyzed in a time-varying fashion (rolling values). As shown, distributional properties of virtual scenarios present a range of values similar to that of our original dataset, but a little more concentrated in smaller values for both metrics. While kurtosis and skewness distributions show more variance between simulations when computed for the whole time series, less differences arise when computed as time-varying features.\par

\begin{figure}[h!]
\begin{center}
\begin{minipage}{\textwidth}
\subfigure[Returns.]{
\resizebox*{0.5\textwidth}{!}{\includegraphics{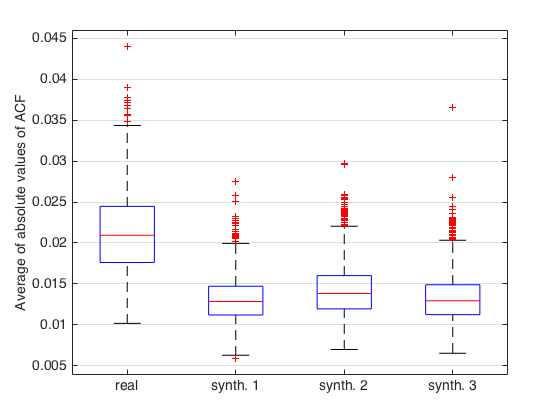}}\label{fig:autocorr_sims}}
\subfigure[Absolute returns.]{
\resizebox*{0.5\textwidth}{!}{\includegraphics{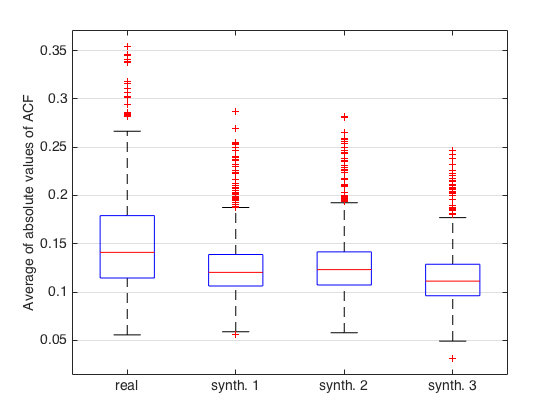}}\label{fig:abs_sims}}
\end{minipage}
\caption{Distribution of the average of absolute values of the ACF of returns (a) and absolute returns (b).
\label{fig:corrs_sims}}
\end{center}
\end{figure}

Regarding dependence properties, Figure \ref{fig:corrs_sims} shows the comparison of autocorrelation values for both returns and absolute returns, as computed in Section \ref{sssec:dep}. As it can be seen in Figure \ref{fig:autocorr_sims}, while simulated returns show the desired absence of autocorrelation, the average of the absolute values of the ACF is significantly lower in the virtual scenarios compared to our original dataset. This can be due to the fact that our returns generation process do not follow an autoregressive approach, and so the ACF at any lag is closer to zero than that of real time series for the first lags. On the other hand, autocorrelation of absolute returns for the virtual scenarios are much closer to that of our original dataset (Figure \ref{fig:abs_sims}), although a bit lower.\par

\begin{figure}[h!]
\begin{center}
\begin{minipage}{\textwidth}
\subfigure[Distributions of trend ratios.]{
\resizebox*{0.5\textwidth}{!}{\includegraphics{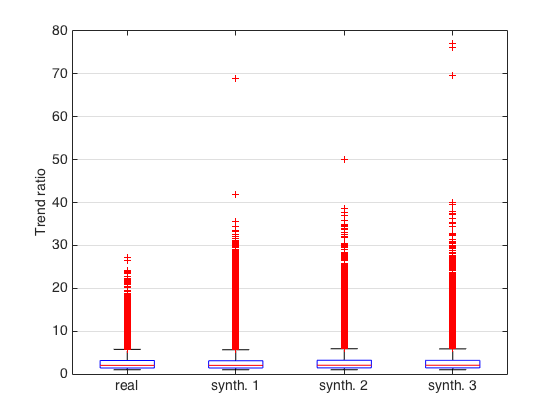}}\label{fig:trend_sims}}
\subfigure[Distributions of directional similarity.]{
\resizebox*{0.5\textwidth}{!}{\includegraphics{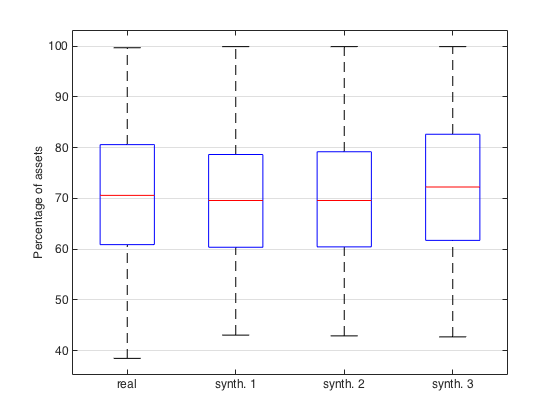}}\label{fig:disim_sims}}
\end{minipage}
\caption{Distributions of trend ratios (a) and directional similarity (b) for the real dataset and the three generated virtual scenarios.
\label{fig:other_sims}}
\end{center}
\end{figure}

In order to summarize the pathwise properties of our artificially generated datasets, Figure \ref{fig:trend_sims} shows the distributions of trend ratios for the three virtual scenarios along with that of the original dataset. As shown, the distributions of simulated data are quite similar to those of the real data but some outliers with high trend ratios may appear. However, as any number of additional assets can be generated through the PCA projection-recovery process, these outliers can be easily avoided by detecting them and discarding the assets in which they appear, generating new ones. Finally, as can be seen in Figure \ref{fig:disim_sims}, similar distributions of directional similarity values are obtained for both real and artificial data, showing the fact that assets in virtual scenarios alternate between high and low correlations along time as real assets do.\par

\section{Conclusions and future work}
\label{sec:conclusions}

In this work we have presented an approach to simulate virtual scenarios of multivariate financial data as long as desired. This approach generates artificial asset returns that behave much like the real ones do, as measured by our sanity-check constraints. Virtual scenarios can be simulated, at a low computational cost, involving decades of trading days for hundreds of assets within a given market, and even new artificial assets for that market can be created.\par

First, in order to define the constraints that artificial asset returns/prices must comply with, the best-known stylized facts have been described and some other properties that account for the cross-asset relationship within a specific market have been introduced. Then, some of the most common approaches found in publicly available toolboxes have been tested in order to check how the simulated asset returns/prices reproduce the observed properties. Among the distributional properties of asset returns, the excess kurtosis is one of the most difficult features to reproduce, specially when it is analyzed as a time-varying feature, even if heavy-tailed generative distributions are used. Neither excess kurtosis or volatility clustering are properly reproduced by stochastic models based on the Geometric Brownian Motion, and volatility models (such as the GARCH family) with Student's t innovations are required to obtain asset returns that better satisfy these constraints. However, when distributional properties of return time series are analyzed as a time-varying feature, more differences arise with real assets. Moreover, none of the available multivariate approaches found as ready-to-use software packages allow to generate virtual scenarios for high-dimensional datasets as the one proposed in this work, involving hundreds or even thousands of assets, because as the number of assets increases, they fail to reproduce volatility clustering and the parameter fitting process fails to converge even for tens of assets.\par

On the other hand, our proposed approach allows to generate financial datasets as large as desired while still reproducing volatility clustering and cross-assets relationships (and their changes over time) to a great extent, leading to sets of assets that behave as belonging to the same market. While pathwise artifacts may appear in the generating process, they can be easily detected and the assets in which they appear replaced by error-free new ones thanks to the PCA projection-recovery process.\par

As it has been shown, a wide variety of scenarios can be generated by simply randomly alternate upward and downward trends, but also handcrafted scenarios could be generated if a trend sequence is specified. Although the behavior of each generated trend in the proposed approach is somewhat constrained as being synthesized from the same sequence of multivariate parameters estimated in the analysis stage, more variability could be introduced by slightly changing some of these parameters. For instance, average returns or covariances may be shifted, or even the length of the trend altered by sampling a number of returns per window in the synthesis stage different to that used in the analysis one. However, further work is required to ensure that the analyzed properties are kept when applying such changes.\par

Regarding distributional properties, kurtosis and skewness values for the simulated assets behave much similar to the real ones than those generated by other analyzed techniques, specially when considered as time-varying features, as the parameters of the sampled distribution change for every short-time window. Finally, while simulated return time series show the desired absence of autocorrelation, some differences appear compared to the real ones as our returns generation process does not follow an autoregressive approach. Although the generated virtual scenarios are still useful for volatility- or trend-driven algorithms, this fact may have some effect on strategies based on autorregressive models, so further research must be done in this direction.\par





\bibliography{sample}



\end{document}